  \documentclass{IEEEtran}
  \usepackage{cite}
  \usepackage{amsmath,amssymb,amsfonts}
  \usepackage{algorithmic}
  \usepackage{graphicx,epstopdf}
  \usepackage{textcomp}
  \usepackage{setspace}
  \usepackage{enumitem}
  \usepackage{slashbox}
  \usepackage[margin=2cm]{geometry}
  \usepackage{adjustbox}
  \usepackage{multirow}
  \usepackage[table]{xcolor}
  \def\BibTeX{{\rm B\kern-.05em{\sc i\kern-.025em b}\kern-.08em
  		T\kern-.1667em\lower.7ex\hbox{E}\kern-.125emX}}
  \usepackage{subfigure}
  \newtheorem{my_theorem}{Theorem}
  
  \newtheorem{my_lemma}{Lemma}
  
  \newtheorem{my_proposition}{Proposition}
  
  \usepackage{float}
  \floatstyle{ruled}
  \newfloat{algorithm}{tbp}{loa}
  \providecommand{\algorithmname}{Algorithm}
  \floatname{algorithm}{\protect\algorithmname}
  \newcommand*{\J}{\jmath}%
    
\title{Performance of Integrated   IoT Network  with Hybrid mmWave/FSO/THz Backhaul Link}
\author{
	\IEEEauthorblockN{Pranay Bhardwaj,~\IEEEmembership{Graduate Student Member,~IEEE}, Vedant Bansal, Nikhil Biyani, Shiv Shukla, and S. M. Zafaruddin, ~\IEEEmembership{Senior Member,~IEEE} }\\

	\thanks{The authors are with the Department of Electrical and Electronics Engineering, Birla Institute of Technology and Science, Pilani, Pilani-333031, Rajasthan, India.  Email:\{p20200026, f20201534, f20200776, f20200413, syed.zafaruddin\}@pilani.bits-pilani.ac.in.}
	
	\thanks{This work was supported	in part by the Science and Engineering Research Board (SERB), Department of Science and Technology (DST), Government of India, through the Mathematical Research Impact Centric Support (MATRICS) scheme under Grant MTR/2021/000890.}
}

\begin{document}
	\maketitle
	
\begin{abstract}		

Establishing  end-to-end connectivity of Internet of Things (IoT) network with the core for collecting sensing data from remote and hard-to-reach terrains  is a challenging task. In this article, we analyze the performance of an IoT network integrated with wireless backhaul link for data collection. 
We propose a solution that involves a self-configuring protocol for aggregate node (AN) selection in an  IoT network, which sends the data packet to an unmanned aerial vehicle (UAV) over radio frequency (RF) channels.  We adopt a novel hybrid transmission technique for wireless backhaul  employing opportunistic selections combining (OSC) and maximal ratio combining (MRC) that simultaneously transmits the data packet on mmWave (mW), free space optical (FSO), and terahertz (THz) technologies to take advantage of their complementary characteristics. We employ the decode-and-forward (DF) protocol to integrate the IoT  and backhaul links and  provide physical layer performance assessment  using outage probability and average bit-error-rate (BER)  under diverse channel conditions. We also develop simplified expressions  to gain a better understanding of the system's performance at high signal-to-noise ratio (SNR). We provide computer simulations to  compare different wireless backhaul technologies under various channel and SNR scenarios
 and demonstrate the performance of the data collection using the integrated link.  

\end{abstract}
	

\section{Introduction}\label{sec:introduction}
The Internet of Things (IoT) has a wide range of applications, spanning from communication and connectivity to healthcare, agriculture, transportation, and more \cite{Fuqaha2015_iot_survey}. IoT devices are capable of collecting critical information in remote and hard-to-reach areas. However, due to the infeasibility of wireline backhaul, there may be connectivity issues between the IoT network and the core network. Moreover, the limited power of IoT devices necessitates consideration of the network's lifetime when connected to the access point (AP) making it challenging to collect data and process it in a timely manner for optimal decision-making.

With the proliferation of connected devices, the process of data collection has become a crucial aspect of IoT networks. Extensive research has been carried out in this area, with numerous studies focusing on various aspects of IoT data collection \cite{Luong2016_iot_data_collection_survey,Xu2018_iot_data_collection,Tao2019_iot_data_collection,Liu2018_iot_data_collection,Wei2022_iot_data_collection,Li2023_iot_data_collection,Pang_2014_data_collection}. Once the data is collected, it needs to be transmitted to a core network for further processing and analysis. Current research primarily focuses on collecting data from IoT devices to an AP or an unmanned aerial vehicle (UAV) with little emphasis on further transmitting it to the core network through a backhaul link (BHL). The algorithms used for data collection must be optimized, considering the energy constraints of the IoT devices. Moreover, the BHL must be capable of handling the amounts of data generated by IoT devices while also providing high reliability. Thus, the development of low-complexity data collection algorithms for IoT networks, coupled with a robust BHL, is essential to ensure seamless connectivity with the core network.

As wireline backhaul installation can be difficult in hard-to-reach areas, recent research has focused on high-bandwidth wireless backhaul solutions \cite{Dehos2014_mW_backhaul,Alzenad208_fso_backhaul,Trinh2017_dual_hop_FSO_RF,Li2019_dual_hop_FSO_RF,Balti2018_dual_hop_rf_fso,Zhang2020_dual_hop_mw_fso,Li2022_dual_hop_thz_fso,Pai2021_dual_hop_THz_backhaul,Li_2021_THz_AF,Bhardwaj2022_systems_journal,Chen2016_hybrid_rf_fso,Makki2016_hybrid_fso_RF,Dahrouj2015_rf_fso_hybrid_backhaul,Enayati2016_hybrid_fso_RF_backhaul,Eryani2018_hybrid_rf_fso,Sharma2019_hybrid_rf_fso,Badarneh2020_mW_FSO_simultaneous,Singya2022_hybrid_fso_thz_backhaul}. Initial Studies have primarily focused on single-technology, high-bandwidth options, such as millimeter-wave (mW), free space optical (FSO), or terahertz (THz) links as the wireless backhaul link \cite{Dehos2014_mW_backhaul,Alzenad208_fso_backhaul,Trinh2017_dual_hop_FSO_RF,Li2019_dual_hop_FSO_RF,Balti2018_dual_hop_rf_fso,Zhang2020_dual_hop_mw_fso,Li2022_dual_hop_thz_fso,Pai2021_dual_hop_THz_backhaul,Li_2021_THz_AF,Bhardwaj2022_systems_journal}. However, recent studies have explored the advantages of hybrid networks, which use multiple distinct technologies and transmit simultaneously to increase link capacity and reliability of the communication link. Several studies have investigated the performance of dual-technology hybrid RF-FSO, mW-FSO, and FSO-THz networks for the BHL \cite{Chen2016_hybrid_rf_fso,Makki2016_hybrid_fso_RF,Dahrouj2015_rf_fso_hybrid_backhaul,Enayati2016_hybrid_fso_RF_backhaul,Eryani2018_hybrid_rf_fso,Sharma2019_hybrid_rf_fso,Badarneh2020_mW_FSO_simultaneous,Singya2022_hybrid_fso_thz_backhaul}. Further, utilizing mixed communication networks consisting of backhaul and access  is a promising method for boosting wireless network capacity and expanding coverage while maintaining low power requirements \cite{Hasna_2004_AF}. This approach involves routing signals from a source to a destination through an intermediate relay node, providing connectivity in scenarios where traditional direct transmission between the source and destination experiences high path-loss and deep fade. The BHLs are generally integrated with the access links using decode-and-forward (DF) or amplify-and-forward (AF) relaying protocol \cite{Trinh2017_dual_hop_FSO_RF,Li2019_dual_hop_FSO_RF,Balti2018_dual_hop_rf_fso,Zhang2020_dual_hop_mw_fso,Li2022_dual_hop_thz_fso,Pai2021_dual_hop_THz_backhaul,Li_2021_THz_AF,Bhardwaj2022_systems_journal,Sharma2019_hybrid_rf_fso,Singya2022_hybrid_fso_thz_backhaul}.

While utilizing two technologies is beneficial, incorporating all three (mW, FSO, and THz) could further improve the quality of service for wireless backhaul. The backhaul technology relying on mmWave/FSO may experience limitations during unfavorable weather conditions since FSO is vulnerable to atmospheric factors like turbulence, fog, and snow. THz transmission possesses distinct qualities: it is susceptible to weather conditions but also subject to increased path loss caused by absorption. As a result, FSO and THz exhibit complementary characteristics. 	The adoption of a triple-technology hybrid backhaul represents a significant advancement with multiple advantages. Integrating an additional link introduces system diversity, leading to enhanced performance, improved reliability, and superior quality of service (QoS) across various scenarios. By harnessing the distinctive characteristics and capabilities of each technology (mmWave, FSO, and THz), the triple-technology backhaul enables a higher QoS. To the best of authors knowledge, no analysis of triple-technology hybrid wireless backhaul mixed with an access link exists in the literature, making it desirable for catering to high data rate and high bandwidth applications in next-generation wireless communication networks. In this paper, we study 	end-to-end performance of data collection from an IoT network in hard-to-access terrains transferred to the core network via an intermediate UAV and a high-speed wireless BHL. We consider  triple-technology in the backhaul to harness the diversity of three  links without increasing the data rate of the backhaul. The major contributions are as follows:
\begin{itemize}	
	\item We develop a self-configuring protocol for aggregate-node (AN) selection, which sends a data packet collected from IoT network to an UAV. Selecting a single AN requires local communication from each device to the AN, thereby eliminating independent transmission from each device to the UAV.
	 \item We utilize a hybrid triple-technology wireless BHL, which transmits the information from the UAV to the core network using  mW, FSO, and THz links simultaneously for high-speed and reliable data transmission. The inclusion of THz connectivity into the current hybrid mmWave/FSO system brings about system diversity, resulting in improved performance, increased reliability, and superior quality of service in different scenarios, particularly in challenging weather conditions.
	\item  We develop statistical characterization of outdoor THz channel considering the mixture Gaussian distribution for the short-term fading and statistical model for pointing errors at THz frequency for aerial communications. This  approach effectively captures the dynamics of UAV-assisted THz transmission, encompassing both channel fading and pointing errors.  Existing literature provides statistics of the  THz link over $\alpha$-$\mu$ fading (a model for indoor scenarios) and  FSO pointing errors.
	\item  We derive the PDF and CDF of the triple-technology BHL incorporating maximal ratio combining (MRC) and opportunistic selection combining (OSC) diversity schemes. We also integrate  the IoT and BHL employing the DF protocol at the UAV and analyze the physical layer performance using the outage probability and average BER.
	
	\item We develop an asymptotic expression for the outage probability of the integrated system and obtain the diversity order to gain insights into the system's performance with change in the system parameters. Computing the simplified asymptotic expression involving the Gamma function allows for a more efficient evaluation of performance at high SNRs.
\end{itemize}

\subsection{Related Works}

Data collection in IoT networks can be challenging due to data volume, difficult terrain, energy limitations of the IoT devices, etc. To ensure effective data collection, IoT networks must have a reliable and robust architecture. In recent years, researchers have focused on addressing the issue of efficient data collection in IoT networks  \cite{Luong2016_iot_data_collection_survey,Xu2018_iot_data_collection,Tao2019_iot_data_collection,Liu2018_iot_data_collection,Wei2022_iot_data_collection,Li2023_iot_data_collection,Pang_2014_data_collection}. The authors in \cite{Xu2018_iot_data_collection} propose an information-centric data collection algorithm which supports the collection of multimedia sensor data in the short range between mobile devices and wireless sensors based on their moving speeds. The work presented in \cite{Tao2019_iot_data_collection} focuses on a secure data collection scheme for an IoT-based healthcare system using field-programmable gate array (FPGA) hardware-based ciphers and secret cipher shared algorithms. In \cite{Wei2022_iot_data_collection}, the authors have presented a thorough survey of UAV assisted data collection technologies including clustering of sensor nodes, UAV data collection mode, and resource allocation. The authors in \cite{Li2023_iot_data_collection} have addressed to the collection maximization issue using UAV involving energy-constraints in an IoT network. The authors in \cite{Pang_2014_data_collection} investigate the use of UAVs to efficiently collect sensed data in wireless rechargeable sensor clusters located in challenging terrains.

Recent research investigates various high-speed hybrid wireless backhaul network as potential alternatives to wireline optical fiber and digital subscriber lines (DSL)  \cite{Chen2016_hybrid_rf_fso,Makki2016_hybrid_fso_RF,Dahrouj2015_rf_fso_hybrid_backhaul,Enayati2016_hybrid_fso_RF_backhaul,Eryani2018_hybrid_rf_fso,Sharma2019_hybrid_rf_fso,Badarneh2020_mW_FSO_simultaneous,Singya2022_hybrid_fso_thz_backhaul}. In \cite{Chen2016_hybrid_rf_fso}, the authors have analyzed a point to multi-point communication system for multiple FSO and RF users using hybrid FSO-RF link. Another study \cite{Dahrouj2015_rf_fso_hybrid_backhaul} presented a low-cost hybrid RF/FSO backhaul solution, where base stations are connected using either optical fiber or hybrid links. In \cite{Enayati2016_hybrid_fso_RF_backhaul}, the authors considered a hybrid FSO/RF backhaul network to provide seamless connectivity in rural areas. The work in \cite{Eryani2018_hybrid_rf_fso} focused on a multiuser mixed RF and hybrid FSO/RF system, where multiple mobile users transmit their data to an intermediate DF relay node through a virtual MIMO RF link. In \cite{Sharma2019_hybrid_rf_fso}, a switching scheme was proposed for a DF-relaying-based hybrid FSO/RF system using MRC at the destination. A study on the performance of a hybrid mmWave/FSO system was described in \cite{Badarneh2020_mW_FSO_simultaneous} and supported by Monte Carlo simulations. In \cite{Singya2022_hybrid_fso_thz_backhaul}, the authors proposed a hybrid FSO/THz-based backhaul network to deliver high data rates to terrestrial mobile users through mmWave access lines. However, none of these studies have investigated the combination of all three communication methods (mmWave, FSO, and THz) simultaneously, which would be highly beneficial for high data rate and high bandwidth applications in next-generation wireless communication networks.

\begin{figure*} [tp]	
 \hspace{2.6cm}	\includegraphics[scale=0.30]{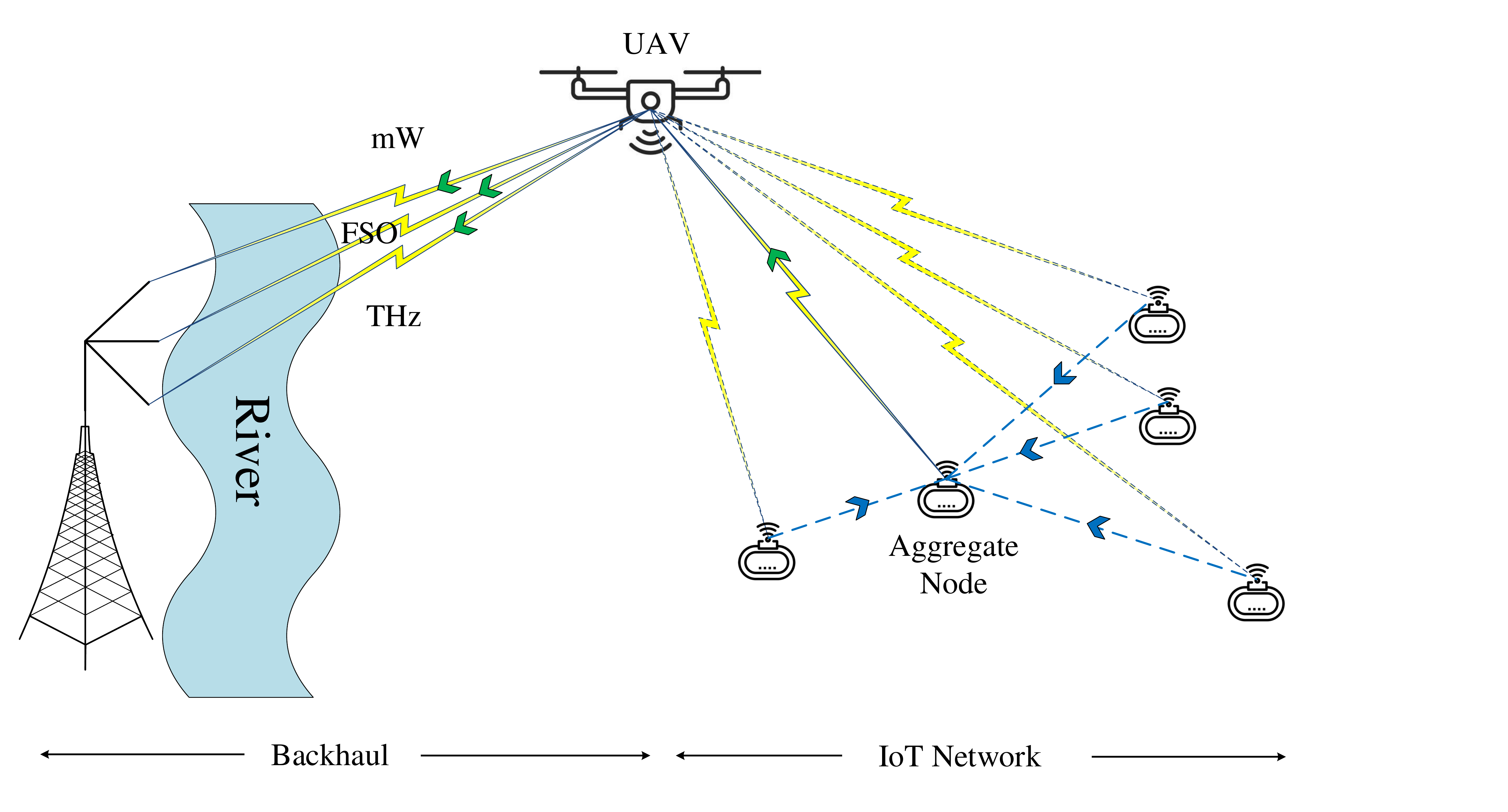}
	\caption{System model consisting of data collection from IoT network and triple-technology hybrid BHL.}
		\label{fig:system_model}
\end{figure*}

Mixed wireless communication has attracted significant attention from researchers due to its ability to increase the range and improve the reliability of communication networks \cite{Trinh2017_dual_hop_FSO_RF,Li2019_dual_hop_FSO_RF,Balti2018_dual_hop_rf_fso,Zhang2020_dual_hop_mw_fso,Li2022_dual_hop_thz_fso,Pai2021_dual_hop_THz_backhaul,Li_2021_THz_AF,Bhardwaj2022_systems_journal,Sharma2019_hybrid_rf_fso,Singya2022_hybrid_fso_thz_backhaul}. In \cite{Trinh2017_dual_hop_FSO_RF}, the authors analyze the outage likelihood, Average BER performance, and ergodic capacity of mixed mmWave RF-FSO system over Rician and $M$-fading channels. In \cite{Balti2018_dual_hop_rf_fso}, the authors have presented a dual-hop mixed RF-FSO system with several relays using AF relaying protocol to address the impact of co-channel interference employing channel-state information (CSI)-assisted AF relay and partial relay selection with outdated CSI by deriving statistical results and performance analysis. In \cite{Zhang2020_dual_hop_mw_fso}, the authors present novel closed-form performance metrics for dual-hop mixed FSO-mW system utilizing AF relaying protocol by considering FTR fading for the mmWave and Gamma-Gamma turbulence for the FSO link. The effectiveness of a mixed THz-FSO system has been analytically examined in \cite{Li2022_dual_hop_thz_fso} using semi-blind AF relaying by considering $\alpha$-$\mu$ fading for THz link and DGG fading for FSO link combined with pointing errors. The authors have also taken taken into account the hardware impairments to derive the outage probability of the hybrid system. The authors in \cite{Pai2021_dual_hop_THz_backhaul,Li_2021_THz_AF} have considered the THz link in the backhaul and the access link is integrated using the AF relaying protocol. In\cite{Bhardwaj2022_systems_journal} the authors have considered an THz link as the backhaul have used the AF relay to mix the multi-antenna RF access link. The authors in \cite{Sharma2019_hybrid_rf_fso} mix the hybrid FSO/RF backhaul with the access link using a selective DF protocol, where the transmission is conditioned based on successful decoding at the relay whereas in \cite{Singya2022_hybrid_fso_thz_backhaul}, the hybrid FSO/THz backhaul is mixed with the access link using DF relaying protocol.

Based on the preceding discussion, it is necessary to develop efficient algorithms for data collection in challenging terrain for IoT networks. Additionally, ensuring the reliable transmission of data from the UAV to the core network is crucial for further information processing, necessitating a robust backhaul link.

\emph{Notations}: The parameter of the $i$-th RF link from the IoT device to the UAV is denoted by $(\cdot)_i$. The Gamma function is represented by $\Gamma(a)= \int_{0}^{\infty}t^{a-1} e^{-t}dt$, while the Meijer's G-function is denoted by $G_{p,q}^{m,n}\big(.|.\big)$. Similarly, the Fox's H-function is denoted by $H_{p,q}^{m,n}\big(.|.\big)$ \cite{Mathai_2010}. The imaginary number is represented by $\J$.

\section{System Description and Channel Models}

The system model illustrated in Fig. \ref{fig:system_model} shows a network consisting of $n$ IoT devices located in a hard-to-reach terrain. These devices are responsible for transmitting and receiving critical sensing data to and from a backhaul network through an UAV. In challenging terrain, it may be impractical or difficult to install traditional base stations due to physical constraints or geographical limitations. The use of UAVs as relays between the IoT devices and the core network offers several advantages such as mobility, flexibility, rapid deployment capabilities, coverage extension potential.

Instead of transmitting the data from every IoT device directly to the UAV,   the other devices in the network send their data to the $n^*$ device, termed as the AN (selected through a self-configuring protocol discussed in the next section) with an SNR $\gamma_{n}^{\rm local}$ at the AN, where $n= 1, 2, 3,\cdots N$ excluding the AN device $n^{*}$. The AN further transmits the collected data from all the other devices to the UAV for an efficient transmission. We define the signal-to-noise ratio (SNR) for the $i$-th IoT device at the UAV as $\gamma_i={\bar{\gamma}}^{\rm IoT}|h_{f}|^2$, where $h_f$ denotes the channel gain coefficient and ${\bar{\gamma}}^{\rm IoT}$ denotes the average SNR. Assuming Rayleigh fading,  $\gamma_i$ is random variable  with  the PDF $f_{\gamma_i}^{\rm IoT}(\gamma) = \frac{1}{{\bar{\gamma}}^{\rm IoT}} {\rm exp}({-\frac{\gamma}{{\bar{\gamma}}^{\rm IoT}}})$ and CDF $F_{\gamma_i}^{\rm IoT}(\gamma) = 1-e^{-\lambda \frac{\gamma}{{\bar{\gamma}}^{\rm IoT}}}$.

We employ a DF-assisted UAV to integrate the IoT link with a wireless backhaul. We use a hybrid transmission approach that employs three high-bandwidth wireless technologies, namely mW, FSO, and THz, simultaneously from the UAV to the core network. 

In the following, we describe the fading channel models  of the BHL involving mW, FSO, and THz. We define the SNR for mW, FSO, and THz channels as  $\gamma^{\rm mW}={\bar{\gamma}^{\rm mW}}|h_{fp}^{\rm FTR}|^2$, $\gamma^{\rm FSO}={\bar{\gamma}^{\rm FSO}}|h_{f}^{\rm {\cal F}}|^2$, and $\gamma^{\rm THz}={\bar{\gamma}^{\rm THz}}|h_{fp}^{\rm { THz}}|^2$,  where $\bar{\gamma}^{\rm mW}, \bar{\gamma}^{\rm FSO}$, and $\bar{\gamma}^{\rm THz}$ denotes the average SNR of the respective transmission technologies.

To model the small-scale fading in the mmWave band, we use the fluctuating two-ray (FTR) fading channel. The FTR channel model is specifically designed to capture the statistical properties of a received signal that consists of dominant specular components along with random fluctuations. It has been shown to provide a more accurate representation of small-scale fading measurements in mmWave communications compared to other fading models \cite{Jiayi2018_FTR}. The PDF and CDF of the FTR  fading channel can be expressed as follows :
\begin{flalign} \label{eq:pdf_ftr_meijer}
	{f_\gamma^{\rm mW}}(\gamma) = & \frac{{{m^m}}}{{\Gamma (m)}}\sum \limits_{j = 0}^\infty {\frac{{K^jd_j \gamma^j}}{{(\Gamma (j + 1))^2(2\zeta ^2)^{j+1}} ({{\bar{\gamma}}^{\rm mW}})^{j+1}}} \nonumber \\ \times & G_{0,1}^{1,0}\bigg(\frac{\gamma}{2\zeta ^2 {\bar{\gamma}}^{\rm mW}}\bigg|\begin{matrix} - \\ 0\end{matrix}\bigg)
\end{flalign}

\begin{flalign} \label{eq:cdf_ftr_meijer}
	{F_{\gamma}^{\rm mW}}(\gamma) = & \frac{{{m^m}}}{{\Gamma (m)}}\sum \limits _{j = 0}^\infty {\frac{{K^jd_j \gamma^j}}{{(\Gamma (j + 1))^2}({\bar{\gamma}}^{\rm mW})^j}} \nonumber \\ \times &  G_{1,2}^{1,1}\bigg(\frac{\gamma}{2 \zeta^2 {\bar{\gamma}}^{\rm mW}}\bigg| \begin{matrix} 1 \\ 1+j, 0	\end{matrix}\bigg)    
\end{flalign}
where $K$ denote the ratio between the average power of the dominant component and that of the multi-path, $m$ represent the index of fading severity, and $\Delta$ signify the similarity between the two dominant waves. The term $\zeta^2$ corresponds to the variance of diffused components, which satisfies $\zeta^2=\frac{1}{2(1+K)}$ for the normalized averaged SNR. Additionally, the factor $d_j$, initially defined in \cite{Jiayi2018_FTR} and updated in \cite{Miguel2021} with an additional factor.

The misalignment of antennas between the IoT device and the UAV, as well as the disorientation of the UAV, can lead to pointing errors in the FSO link. Thus, the PDF of the pointing error is given by \cite{Najafi2020_uav_pointing} \cite{Boulogeorgos_PE_2022}:
\begin{flalign} \label{eq:poinitng_fso}
	f_{h_p}(h_p) = \frac{\phi}{S_0} \left(\frac{h_p}{S_0}\right)^{\phi-1}; ~~~~ 0 \leq h_p \leq S_0
\end{flalign}

Here, $S_0$ is defined as ${\rm erf}(v_{\rm min}) {\rm erf}(v_{\rm max})$, with $v_i = \frac{\zeta}{w(L_2)} \sqrt{\frac{\pi}{2 \chi_i}}$ for $i \in \{{\rm min}, {\rm max}\}$. We define $\chi_{\rm min} = \frac{2}{\chi_y + \chi_z + \sqrt{(\chi_y-\chi_z)^2 + 4 \chi_{yz}^2}}$ and $\chi_{\rm max} = \frac{2}{\chi_y + \chi_z - \sqrt{(\chi_y-\chi_z)^2 + 4 \chi_{yz}^2}}$, where $\chi_y = \cos^2(\theta) + \sin^2(\theta) \cos^2(\delta)$, $\chi_z = \sin^2(\theta)$, and $\chi_{yz} = -\cos(\theta)\sin(\theta)\sin(\delta)$ with $\theta$ and $\delta$ are defined in \cite{Najafi2020_uav_pointing}. The radius of the effective area of the UAV antenna is denoted by $\zeta$. The beam-width at distance $L_2$, given by $w(L_2)$, is given by $w(L_2) = w_0 \sqrt{1+ \left(1+\frac{2w_0^2}{\chi^2(L_2)}\right) \left(\frac{cL_2}{\pi f w_0^2}\right)^2}$, where $w_0$ represents the beam-waist radius, $c$ is the speed of light, $f$ is the transmission frequency, and $\chi(L_2) = (0.55 C_n^2 k^2 L_2)^{-\frac{3}{5}}$. Here, $C_n^2$ denotes the index of the refraction structure parameter, and $k = \frac{2\pi f}{c}$ represents the wave number. The value of $\phi$ in \eqref{eq:poinitng_fso} is defined as $\phi = \frac{k_m(w(L_2))^2}{4 \sigma_p^2 + 4 d_x^2\sigma_0^2}$, where $k_m = \frac{k_{\rm min}+k_{\rm max}}{2}$ and $k_i = \frac{\sqrt{\pi} \chi_i {\rm erf}(v_i)}{2 v_i \exp(-v_i^2)}$ for $i \in {{\rm min}, {\rm max}}$.

We utilize ${\cal{F}}$-turbulence channel to model the turbulence in the FSO link. The ${\cal{F}}$-turbulence channel is a suitable choice for modeling FSO wireless channels that exhibit both weak and strong irradiance fluctuations. It offers a more accurate fit to experimental and computer simulation data compared to other models. Additionally, the density and distribution functions of the ${\cal{F}}$-turbulence channel can be represented in a simplified form, making it a convenient for modeling and analysis purposes\cite{Badarneh2020_mW_FSO_simultaneous}. The PDF and CDF of the FSO channel experiencing turbulence in presence of the the zero-boresight pointing error are given by \cite{Badarneh2020_mW_FSO_simultaneous}
\begin{equation}
	f_{\gamma}^{\rm FSO}(\gamma)=\frac{\alpha \phi \gamma^{-\frac{1}{2}} G_{2,2}^{2,1}\left[\frac{\alpha \gamma^{\frac{1}{2}}}{\psi ({\bar{\gamma}}^{\rm FSO})^{\frac{1}{2}}}\left|\begin{array}{c} -\beta,\phi \\ \beta-1,\phi-1\\ \end{array}\right.\right]}{({\bar{\gamma}}^{\rm FSO})^{\frac{1}{2}} \psi\Gamma(\alpha)\Gamma(\beta)}
	\label{eq:PDF_F_pe}
\end{equation}
\begin{flalign} \label{eq:CDF_F_pe}
	F_{\gamma}^{\rm FSO}(\gamma)= \frac{\alpha_{{\scriptscriptstyle  F}}\phi \gamma^{\frac{1}{2}} G_{3,3}^{2,2}\left[\frac{\alpha \gamma^{\frac{1}{2}}}{\psi ({\bar{\gamma}}^{\rm FSO})^{\frac{1}{2}}}\left|\begin{array}{c}	-\beta, 0, \phi \\ \beta-1,\phi-1, -1 \end{array}\right.\right]}{{({\bar{\gamma}}^{\rm FSO})}^{\frac{1}{2}} \psi\Gamma(\alpha)\Gamma(\beta) }	
\end{flalign}
where $\alpha=\frac{1}{\exp(\sigma_{\ln \scriptscriptstyle S}^{2})-1}$, $\beta=\frac{1}{\exp(\sigma_{\ln \scriptscriptstyle \mathcal{L}}^{2})-1}$, and  $\psi = (\beta-1) S_0$. The terms $\sigma_{\ln \scriptscriptstyle S}^{2}$ and $\sigma_{\ln \scriptscriptstyle \mathcal{L}}^{2}$ represent the variances of small and large scale log-irradiance, respectively \cite{Badarneh2020_mW_FSO_simultaneous}.

For the THz link, recently the authors in \cite{Papasotiriou2023_scintific_report_outdoor} have demonstrated through experiments that short-term fading in outdoor THz environments can be precisely modeled using a mixture Gaussian fading model. The PDF of the mixture Gaussian distribution is given by:
\begin{flalign} \label{eq:pdf_channel_thz_gm}
	f_{|h_{f}|}^{\rm THz}(x) = \sum_{i=1}^{K} w_i \frac{\exp\Big(-\frac{(x-\mu_i)^2}{2 \sigma_i^2}\Big)}{\sqrt{2\pi}\sigma_i}
\end{flalign}
where the mean and standard deviation of the $i$-th Gaussian mixture (GM) component are denoted by $\mu_i$ and $\sigma_i$, respectively. The number of GM components is represented by $K$, and the weight of the $i$-th mixture component is given by $w_i \in [0,1]$ such that $\sum_{i=1}^{K} w_i=1$.

Previous research studies have used the FSO pointing error model for the THz band to analyze performance. We use a new pointing error model specifically designed for the THz band as proposed by the authors in \cite{Badarneh2023_THz_Pointing}. The model is suitable for aerial mobile communication, with the PDF given as:
\begin{flalign} \label{eq:pdf_pointing_thz_gm}
	f_{h_{p}}^{\rm THz}(y) = -\rho^2 \ln(y)  y^{\rho-1}
\end{flalign}
where $0<y<1$ and $\rho$ is the pointing error parameter. Higher values of $\rho$ corresponds to the lower pointing errors and vice versa.

\begin{algorithm}[tp]
	\caption{{\color{black}Distributed Data Collection from IoT Network}}
		\textbf{Initialization:}
	\begin{enumerate}
		\item \textbf{PDF Estimation}---Find the PDF of SNR $f_\gamma^{\rm IoT}(\gamma) = \frac{1}{{\bar{\gamma}}^{\rm IoT}} e^{-\frac{\gamma}{{\bar{\gamma}}^{\rm IoT}}}$ by fitting  instantaneous  SNR  from IoT devices to the UAV.
		\item \textbf{Coherence Time}---  Estimate the coherence time $t^{\rm IoT}$ of the channel between the UAV and IoT devices.
		\item  \textbf{Throughput Estimation}---  Estimate the throughput of the transmission as $R=\min(R^{\rm BHL}, R^{\rm IoT})$, where $R^{\rm BHL}$, and $R^{\rm IoT}$ is the throughput of BHL and IoT network, respectively.
		\item  \textbf{Data Packet Size}---  Estimate the data packet size for the $i$-th device as $L_i=(R\times t^{\rm IoT})/N $ such that the packet size of the AN becomes $L=NL_i$,
		
	\end{enumerate}

	\textbf{For each new packet $L$} do
	
	\textbf{A. Selection of AN} --- For each IoT device $i=1$ to $N$ 
	\begin{enumerate}
		\item The $i$-th device measures its own SNR $\gamma_i$ and computes back-off time $\tau_i$ by mapping $\tau_i= f_\gamma^{\rm IoT}(\gamma_i)$
		\item The device with minimum $\tau_i<\tau_j, \forall i,j, i\neq j$ communication with all other devices announcing being the AN.
		\item All the devices send their sensing information to the AN.
	\end{enumerate}
	\textbf{B. Transmission} ---
	\begin{enumerate}
		\item The  AN  transmits the training data to the UAV. The UAV estimates the channel using the training data.
		\item The AN transmits the information data to the UAV
		\item The UAV applies the DF protocol to decode the data.
		\item The UAV forwards the data to the backhaul link using mW, FSO, and THz technologies by applying a frequency up-converter for mW and THz and RF to optical converted for 	FSO.
	\end{enumerate}
	\textbf{B. Data Collection} ---
	\begin{enumerate}
		\item The received signals from the BHL is received at a base station.
		\item The base station  applies either MRC or OSC to decode the data from the UAV.
		\item The decoded data is sent to the central unit for post-processing.
	\end{enumerate}
	\textbf{End}
\end{algorithm}

\section{Protocol Description for Aggregate-Node Selection}
Consider a scenario where an IoT network consists of $N$ devices, and each device has an independent data packet that needs to be transmitted to a core network.  To facilitate this transmission, we employ an UAV as a relay. There are two potential approaches for transmitting data from each device to the UAV: (i) direct transmission from each device to the UAV, or (ii) the devices send their data to a chosen AN, which then transmits the aggregated data to the UAV. We opt for the second method, wherein $N$-$1$ devices locally transfer their data to the AN selected to minimize power consumption  and reduce transmission overhead.  
	
The choice of AN can be made using either a centralized procedure, in which the UAV measures the SNR of each device and selects the device with the highest SNR as the AN, or a distributed approach, where each device undergoes a self-configuring process to become an AN. In order to simplify the centralized control complexity at the UAV, we employ a distributed algorithm to select an AN from a set of $N$ IoT devices  \cite{Bletsas2006}. In our particular study, we assume that each IoT device measures its own SNR and has the opportunity to become the AN based on a back-off time. A common function based on the PDF of the SNR  $f_\gamma^{\rm IoT}(\gamma) = \frac{1}{{\bar{\gamma}}^{\rm IoT}} e^{-\frac{\gamma}{{\bar{\gamma}}^{\rm IoT}}}$ is taken to compute the back-off time   $\tau_i$ by mapping $\tau_i= f_\gamma^{\rm IoT}(\gamma_i)$, where $\gamma_i$ is the SNR of the $i$-th device. The device with the minimum $\tau_i<\tau_j, \forall i,j, i\neq j$ establishes communication with all other devices by announcing itself as the AN over a control channel. Note that the distributed protocol introduces latency corresponding to the back-off time of the AN device.

Subsequently, other IoT devices share their respective data packets with the AN at an SNR $\gamma_n^{\rm local}$, where $n=1, 2, 3, \cdots, N$ excluding the AN device index. The AN transmits the combined data from all the devices to the UAV. Upon receiving the data, the UAV applies the DF protocol and forwards the data to the backhaul link using mW, FSO, and THz technologies by applying a frequency up-converter for mW and THz and RF to optical converted for 	FSO.  A schematic implementation approach of the protocol is  described in Algorithm 1.
		
The average communication resources (such as transmit power) needed for transmitting data from IoT devices to the AN through local communication will be significantly lower compared to the direct transmission from each device to the UAV at a similar quality of service (QoS). Thus, we assume $\gamma_{n}^{\rm local}>>\gamma_i$  such that the performance is limited by the link between the AN (selected from $N$ IoT devices)  and UAV.

It is important to consider that the proposed approach of utilizing multiple high-frequency bands in the BHL comes with the trade-off of improved performance. While it offers potential benefits in terms of increased reliability and QoS, it also introduces the challenge of managing the computational demands required for signal processing.

\section{Statistical Results for BHL}
In this section, we derive the PDF and CDF for the hybrid BHL using MRC and OSC diversity schemes. To obtain the statistical analysis of the hybrid BHL, first, we require the PDF and CDF of the SNR of the outdoor THz channel, combined with the new statistical model for pointing error. The following theorem outlines the derivation of the PDF and CDF for the THz link.

To provide a clear understanding of the derivation process, we outline the steps involved before delving into the specific details. The derivation begins by developing the PDF and CDF for the Gaussian outdoor THz channel model with THz pointing errors by utilizing the Mellin Barnes type integral and generalized Fox's H-function. Next, we proceed to derive the PDF and CDF for the hybrid BHL system employing the MRC and OSC diversity scheme. To obtain the statistical results for MRC, we employ the Moment Generating Function (MGF) approach to calculate the sum of random variables, while for the OSC we utilize the order statistics to determine the signal with the maximum Signal-to-Noise Ratio (SNR).
\begin{my_lemma} \label{th:pdf_cdf_thz_pointing}
	The PDF and CDF of the SNR for the  Gaussian outdoor THz channel combined with the statistical model for THz pointing error is given by 
	\begin{flalign} \label{eq:pdf_thz_combined}
		f_{\gamma}^{\rm THz}(\gamma) =  & \sum_{i=1}^{K} \frac{\rho^2 w_i \exp\Big(\frac{-\mu_i^2}{2 \sigma_i^2}\Big) \gamma^{-\frac{1}{2}}}{2 \sqrt{2\pi}\sigma_i ({\bar{\gamma}}^{\rm THz})^{\frac{1}{2}} } \nonumber \\ \times &   H^{0,2:1,0;1,0}_{2,2:0,1;0,1}\Bigg[\frac{\gamma}{{\bar{\gamma}}^{\rm THz}}, \frac{-2\mu_i \gamma^{\frac{1}{2}}}{({\bar{\gamma}}^{\rm THz})^{\frac{1}{2}}} \Bigg| \begin{matrix} T_1 \\ T_2 \end{matrix}\Bigg] 
	\end{flalign}
	
	\begin{flalign}  \label{eq:cdf_thz_combined}
		F_{\gamma}^{\rm THz}(\gamma) = & \sum_{i=1}^{K} \frac{\rho^2 w_i \exp\Big(\frac{-\mu_i^2}{2 \sigma_i^2}\Big) \gamma^{\frac{1}{2}}}{2 \sqrt{2\pi}\sigma_i ({\bar{\gamma}}^{\rm THz})^{\frac{1}{2}}}  \nonumber \\ \times &   H^{0,3:1,0;1,0}_{3,3:0,1;0,1}\Bigg[\frac{\gamma}{{\bar{\gamma}}^{\rm THz}}, \frac{-2\mu_i \gamma^{\frac{1}{2}}}{({\bar{\gamma}}^{\rm THz})^{\frac{1}{2}}} \Bigg| \begin{matrix} T_3 \\ T_4 \end{matrix}\Bigg] 
	\end{flalign}
	where $T_1 = \bigl\{(\rho-1;2,1)^2\bigr\}:\bigl\{-\bigr\};\bigl\{-\bigr\}$, $T_2 = \bigl\{(\rho-2;2,1)^2\bigr\}:\bigl\{(0,1)\bigr\};\bigl\{(0,1)\bigr\}$, $T_3 = \bigl\{(\rho-1;2,1)^2, (0;2,1)\bigr\}:\bigl\{-\bigr\};\bigl\{-\bigr\}$, and $T_4 = \bigl\{(\rho-2;2,1)^2, (-1;2,1)\bigr\}:\bigl\{(0,1)\bigr\};\bigl\{(0,1)\bigr\}$.	
\end{my_lemma}

\begin{IEEEproof}
	Proof is given in Appendix A.
\end{IEEEproof}

Next, we apply the diversity schemes (MRC and OSC) to improve the performance of the BHL link. For the BHL-MRC, the received signals from the UAV are combined such that the SNR is maximized. Thus, the combiner output SNR is the sum of the SNRs of each wireless link:
\begin{flalign}
\gamma^{\rm MRC} =  \gamma^{\rm mW} + \gamma^{\rm FSO} + \gamma^{\rm THz}
\end{flalign}

Here, we use the moment generation function (MGF) based approach to develop statistical analysis for the MRC system. The PDF of the SNR for the BHL-MRC can be derived using the inverse Laplace transform as
\begin{flalign}
f_{\gamma}^{\rm MRC} = \frac{1}{2\pi \J} \int_{\mathcal{L}} \mathcal{M}_{\gamma}^{\rm MRC} (s) e^{s\gamma} ds
\end{flalign}
where $\mathcal{L}$ denotes the contour, and $\mathcal{M}_{\gamma}^{\rm MRC} (s)$ denotes the MGF of $\gamma^{\rm MRC}$, which is defined as
\begin{flalign} \label{eq:mgf_mrc_eqn}
	\mathcal{M}_{\gamma}^{\rm MRC} (s) = \mathcal{M}_{\gamma}^{\rm mW} (s) \mathcal{M}_{\gamma}^{\rm FSO} (s) \mathcal{M}_{\gamma}^{\rm THz} (s) 
\end{flalign}
with MGF of the individual link given as
\begin{flalign} \label{eq:mgf_eqn}
	 \mathcal{M}_{\gamma}(s) = \int_{0}^{\infty} e^{-s\gamma} f_\gamma(\gamma) d\gamma 
\end{flalign}
 
In the following theorem, we develop the PDF of the SNR for the BHL-MRC:
\begin{my_theorem}
	If the mW, FSO, THz wireless channels and THz pointing errors are distributed according to \eqref{eq:pdf_ftr_meijer}, \eqref{eq:PDF_F_pe}, and \eqref{eq:pdf_thz_combined}, respectively, then the PDF and CDF of the SNR for BHL-MRC is given by	
	\begin{flalign} \label{eq:pdf_hybrid_mrc}
		&f_{\gamma}^{\rm MRC} (\gamma) = \frac{{{m^m}}}{{\Gamma (m)}}\sum \limits_{j = 0}^\infty {\frac{K^jd_j \gamma^{1+j} }{{(\Gamma (j + 1))^2(2\zeta ^2)^{j+1}} ({{\bar{\gamma}}^{\rm mW}})^{j+1}}} \nonumber \\ & \times \sum_{i=1}^{K} \frac{\rho^2 w_i \exp\Big(\frac{-\mu_i^2}{2 \sigma_i^2}\Big) }{2\sqrt{2\pi}\sigma_i ({\bar{\gamma}}^{\rm THz})^{\frac{1}{2}}}  \frac{\alpha\phi}{\psi\Gamma(\alpha)\Gamma(\beta)}   \nonumber \\ &   H^{0,3:1,1;2,2;1,0;1,0}_{3,3:1,1;3,2;0,1;0,1} \left[\begin{matrix} \frac{\gamma}{2\zeta ^2 {\bar{\gamma}}^{\rm mW}}, \frac{\alpha \gamma^{\frac{1}{2}} }{\psi {({\bar{\gamma}}^{\rm FSO})}^{\frac{1}{2}} } ,
			\frac{\gamma}{{\bar{\gamma}}^{\rm THz}} , \frac{-2\mu_i \gamma^{\frac{1}{2}}}{({\bar{\gamma}}^{\rm THz})^{\frac{1}{2}}}  \end{matrix} \Bigg| \begin{matrix} T_3 \\ T_4	\end{matrix}\right]
	\end{flalign}
	
	where $T_3 = \bigl\{(\rho-1;0,0,2,1)^2, (\frac{1}{2};0,0, 1, \frac{1}{2})\bigr\}: \bigl\{(-j,1)\bigr\}; \bigl\{(-\beta,1),(\frac{1}{2},\frac{1}{2}) (\phi,1)\bigr\} ; \bigl\{-\bigr\} ; \bigl\{-\bigr\}   $ and $T_4 = \bigl\{(\rho-2;0,0,2,1)^2, (-j;1,\frac{1}{2},1,\frac{1}{2})\bigr\}: \bigl\{(0,1)\bigr\}; \bigl\{(\beta-1,1), (\phi,1)\bigr\} ; \bigl\{(0,1)\bigr\};\bigl\{(0,1)\bigr\}  $.
	
	\begin{flalign} \label{eq:cdf_hybrid_mrc}
		&F_{\gamma}^{\rm MRC} (\gamma) =  \frac{{{m^m}}}{{\Gamma (m)}}\sum \limits_{j = 0}^\infty {\frac{K^jd_j \gamma^{2+j} }{{(\Gamma (j + 1))^2(2\zeta ^2)^{j+1}} ({{\bar{\gamma}}^{\rm mW}})^{j+1}}} \nonumber \\ & \times \sum_{i=1}^{K} \frac{\rho^2 w_i \exp\Big(\frac{-\mu_i^2}{2 \sigma_i^2}\Big) }{2\sqrt{2\pi}\sigma_i ({\bar{\gamma}}^{\rm THz})^{\frac{1}{2}}}  \frac{\alpha\phi}{\psi\Gamma(\alpha)\Gamma(\beta)}    \nonumber \\ &   H^{0,4:1,1;2,2;1,0;1,0}_{4,4:1,1;3,2;0,1;0,1} \left[\begin{matrix} \frac{\gamma}{2\zeta ^2 {\bar{\gamma}}^{\rm mW}}, \frac{\alpha \gamma^{\frac{1}{2}} }{\psi {({\bar{\gamma}}^{\rm FSO})}^{\frac{1}{2}} } 
			\frac{\gamma}{{\bar{\gamma}}^{\rm THz}} , \frac{-2\mu_i \gamma^{\frac{1}{2}}}{({\bar{\gamma}}^{\rm THz})^{\frac{1}{2}}} \end{matrix} \Bigg| \begin{matrix} T_5 \\ T_6	\end{matrix}\right]
	\end{flalign}
	where $T_5 = \bigl\{(\rho-1;0,0,2,1)^2, (\frac{1}{2}; 0, 0, 1, \frac{1}{2}), (-1-j;1,\frac{1}{2},1,\frac{1}{2})\bigr\}: \bigl\{(-j,1)\bigr\} ; \bigl\{(-\beta,1),(\frac{1}{2},\frac{1}{2}) (\phi,1)\bigr\} ; \bigl\{-\bigr\} ; \\ \bigl\{-\bigr\} ;  $ and $T_6 = \bigl\{(\rho-2;0,0,2,1)^2, (-j;1,\frac{1}{2},1,\frac{1}{2}), (-2-j;1,\frac{1}{2},1,\frac{1}{2})\bigr\}: \bigl\{(0,1)\bigr\} ; \bigl\{(\beta-1,1), (\phi,1)\bigr\} ; \bigl\{(0,1)\bigr\};\bigl\{(0,1)\bigr\} $.
\end{my_theorem}

\begin{IEEEproof}
	See Appendix B.
\end{IEEEproof}

In BHL-OSC, the SNR of each link (i.e., mW, FSO, and THz links) is measured and the link with the maximum SNR value is selected. The resulting SNR in this case will be
\begin{flalign}
\gamma^{\rm OSC} = \max\{ \gamma^{\rm mW}, \gamma^{\rm THz}, \gamma^{\rm FSO}\}
\end{flalign}

Substituting the CDFs of the individual links from \eqref{eq:cdf_ftr_meijer}, \eqref{eq:CDF_F_pe}, and \eqref{eq:cdf_thz_combined} into \eqref{eq:cdf_sc_eqn}, we get the CDF of the BHL-OSC, as compiled in the following proposition:.
\begin{my_proposition}
The CDF is given by \cite{papoulis_2002}
\begin{flalign} \label{eq:cdf_sc_eqn}
F_{\gamma}^{\rm OSC}(\gamma) = F_{\gamma}^{\rm mW}(\gamma) F_{\gamma}^{\rm FSO}(\gamma) F_{\gamma}^{\rm THz}(\gamma)
\end{flalign}
\end{my_proposition}

\begin{IEEEproof}
	The proof is  a straightforward application of the maximum of $3$ random variables.
\end{IEEEproof}

The use of Fox's H representation for statistical analysis is increasingly prevalent in the research community. This allows researchers to obtain insights into the behavior of complex systems and processes, which would have been difficult to obtain otherwise. Moreover, it provides asymptotic expressions in terms of the simpler Gamma function to gain a better understanding of the performance of the system and identifying key parameters, which are essential for performance optimization.

\section{Performance Analysis for Integrated Link}
In this section, we will use the statistical results obtained in the previous section to analyze physical layer performance metrics such as the outage probability and average BER for the considered system.  Additionally, we derive an asymptotic expression for the outage probability in  high SNR region to determine the diversity order of the integrated system for a better engineering insight.

Using the DF relay scheme at the UAV, the resulting SNR of the integrated link can be expressed as follows:
\begin{flalign}\label{eq:IL}
\gamma^{\rm IL} = \min\bigl\{\gamma^{\rm IoT}, \gamma^{\rm BHL}\bigr\}
\end{flalign}
where, the SNR of the AN  at the UAV is given by \cite{papoulis_2002}
\begin{flalign}
\gamma_{}^{\rm IoT} = \max\big\{ {\gamma_1} , {\gamma_2}, \cdots, {\gamma_N}\big\} 
\end{flalign}
where $\gamma_1$, $\gamma_2$, $\cdots$ $\gamma_N$ are individual SNRs of the $N$ IoT devices at the UAV under Rayleigh fading channel. The CDF of the device with maximum SNR is given by \cite{papoulis_2002}
\begin{flalign}
F_{\gamma}^{\rm IoT}(\gamma) = \prod_{i=1}^{N}  F_{\gamma_{\color{blue}i}}(\gamma)
\end{flalign}

Considering identical fading characteristics, the resultant CDF can be given as
\begin{flalign} \label{eq:cdf_iot_eqn}
F_{\gamma}^{\rm IoT}(\gamma)  = (F_{\gamma_i}(\gamma))^N=\Big(1-e^{-\lambda \frac{\gamma}{{\bar{\gamma}}^{\rm IoT}}}\Big)^N
\end{flalign}

We use the binomial expansion as $(x+y)^n$ = $\sum_{k=0}^{n}$ $n \choose k$ $x^{n-k} y^k$ : $(1-x)^n$ = $\sum_{k=0}^{n}$ $n \choose k$ $ (-x)^k$ to get  \eqref{eq:cdf_iot_eqn} as:
\begin{flalign} \label{eq:cdf_iot}
F_{\gamma}^{\rm IoT}(\gamma) = \sum_{k=0}^{N} {N \choose k}  \bigg(-e^{-\frac{\lambda \gamma}{{\bar{\gamma}}^{\rm IoT}}}\bigg)^{k}  
\end{flalign}

\subsection{Outage Probability}
Outage probability refers to the probability that the SNR of a communication system falls below a certain threshold level $\gamma_{\rm th}$, i.e $P(\gamma < \gamma_{\rm th})$. The outage probability of the integrated link is given in the following lemma.
\begin{my_lemma} \label{lemma:outage_relay_eqn}
	The outage probability of the integrated link 	$P_{\rm out}^{\rm IL}$ is given by
	\begin{flalign} \label{eq:outage_relay_eqn}
		P_{\rm out}^{\rm IL} = P_{\rm out}^{\rm IoT} + P_{\rm out}^{\rm BHL} - P_{\rm out}^{\rm IoT} P_{\rm out}^{\rm BHL} 
	\end{flalign}
	where  $P_{\rm out}^{\rm IoT}$ is the outage probability of the IoT network and $P_{\rm out}^{\rm BHL}$ is is the outage probability of the BHL.
	\end{my_lemma}

\begin{IEEEproof}
	Using \eqref{eq:IL}, a simple application of the theory of random variables leads to the  CDF of the integrated link:
	\begin{flalign} \label{eq:CDF_DF}
	F_{\gamma}^{\rm IL} (\gamma) =  F_{\gamma}^{\rm IoT} (\gamma) + F_{\gamma}^{\rm BHL} (\gamma)  -  F_{\gamma}^{\rm IoT} (\gamma) F_{\gamma}^{\rm BHL} (\gamma)  
	\end{flalign}	
	Thus, the outage probability of the IoT link $P_{\rm out}^{\rm IoT}$ can be derived by substituting $\gamma = \gamma_{\rm th}$ in  \eqref{eq:cdf_iot}. To derive $P_{\rm out}^{\rm BHL}$, we substitute  $\gamma = \gamma_{\rm th}$ in the CDF of the BHL, as given in   \eqref{eq:cdf_hybrid_mrc} for MRC or in \eqref{eq:cdf_sc_eqn} for OSC.  Compiling the outage probabilities of individual links,  we get the outage probability of the integrated system  in \eqref{eq:outage_relay_eqn}.
\end{IEEEproof}

\begin{table*}[tp] 
	\centering
	\caption{List of Simulation Parameters} 
		\label{tab:simulation_parameters} 
		\centering 
		\begin{tabular}{|c| c| c| c|} 
			\hline   
			\textbf{Parameter}  & \textbf{Value} &  \textbf{Parameter}  & \textbf{Value} \\ [1ex] 
			\hline  
			Carrier frequency - mmWave  & $ 50 $ \mbox{GHz}  & Transmit power & $0$ \mbox{dBm} \\
			\hline
			Carrier frequency - THz  & $ 300 $ \mbox{GHz}  & Noise Power (mmWave, FSO, THz) &  $ -131 $ \mbox{dBm} \\
			\hline
			Carrier frequency - RF  & $ 800 $ \mbox{MHz}  &  Noise Power (RF) &  $ -144 $ \mbox{dBm} \\
			\hline
			Operating Wavelength - FSO  & $ 750 $ \mbox{nm}  & $\theta$ & $-\frac{\pi}{4}$ \\
			\hline
			Transmitting Antenna Gain (mmWave)  & $ 10 $ \mbox{dBi}& $\delta$ & $\frac{3\pi}{4}$ \\
			\hline
			Receiving Antenna Gain (mmWave) &  $ 26$ \& $27 $ \mbox{dBi} &  $\sigma_0$ & $0.1$ \mbox{rad}  \\
			\hline
			Transmitting Antenna Gain (THz)  & $ 10 $ \mbox{dBi}& $\sigma_p$ & $0.05$ \mbox{rad} \\
			\hline
			Receiving Antenna Gain (THz) &  $ 51 $ \mbox{dBi} & $ d_x $ &  $0.1$ \\
			\hline
			Transmitting Antenna Gain (RF)  & $ 0 $ \mbox{dBi}& $w_0$ & $1$ \mbox{mm} \\		
			\hline
			 Receiving Antenna Gain (RF) &  $ 0 $ \mbox{dBi} & $\rho$ & $27.94$ \\		
			\hline
			 BHL Distance &  $ 500$ \&  $1 $ \mbox{km} & IoT Link Distance &  $ 100 $ \mbox{m} \\		
			\hline
		\end{tabular}
		
\end{table*}


In order to determine the diversity order of the system, we express the outage probability in the high SNR region using asymptotic expressions for both the BHL and IoT links. To obtain the asymptotic outage probability for the BHL-MRC, we  calculate the residue on the dominant poles, resulting in the following expression:
\begin{flalign} \label{eq:asymp_mrc}
	F_{\gamma}^{\rm MRC^\infty} (\gamma) =& C_1 \bigg(\frac{\gamma}{{\bar{\gamma}}^{\rm THz}}\bigg)^{\hspace{-1mm}\frac{1}{2}} \hspace{-1mm} + C_2 \bigg(\frac{\gamma}{{\bar{\gamma}}^{\rm THz}}\bigg)^{\hspace{-1mm}\frac{\rho}{2}} \hspace{-1mm} + C_3 \bigg(\frac{\gamma}{{\bar{\gamma}}^{\rm FSO}}\bigg)^{\hspace{-1mm}\frac{\beta}{2}} \nonumber \\  + & C_4 \bigg(\frac{\gamma}{{\bar{\gamma}}^{\rm FSO}}\bigg)^{\frac{\phi}{2}} + C_5 \bigg(\frac{\gamma}{{\bar{\gamma}}^{\rm mW}}\bigg)^{1}
\end{flalign}
where $C_3$, $C_4$ and $C_5$ are constants. It is evident by observing the exponents of average SNR in \eqref{eq:asymp_mrc} that the diversity order of the BHL-MRC is $\min\Bigl\{1, \frac{\beta}{2}, \frac{\phi}{2}, \frac{1}{2}, \frac{\rho}{2}\Bigr\}$. Similarly, the diversity order for the BHL-OSC can be obtained by individually calculating diversity order for each of the three links in the BHL, which turns  out to be the same as that of the BHL-MRC. The diversity order for the $N$ Rayleigh faded links is equal to $N$. Hence, the  diversity order of the outage probability for the integrated system is given by:
\begin{flalign} \label{eq:diversity_order_total}
	DO^{\rm OP} = \min\Bigl\{1, \frac{\beta}{2}, \frac{\phi}{2}, \frac{1}{2}, \frac{\rho}{2}, N\Bigr\}.
\end{flalign}
The diversity order in  \eqref{eq:diversity_order_total}  depicts varios design and deployment scenarios using various channel and system parameters. It should be noted that the factor $1$ is due to the FTR channel model for the mW propagation.

\begin{figure*}[t]
	\centering
	\subfigure[Outage probability with $ \gamma_{\rm th}=7 \mbox{dB} $]{\includegraphics[scale=0.35]{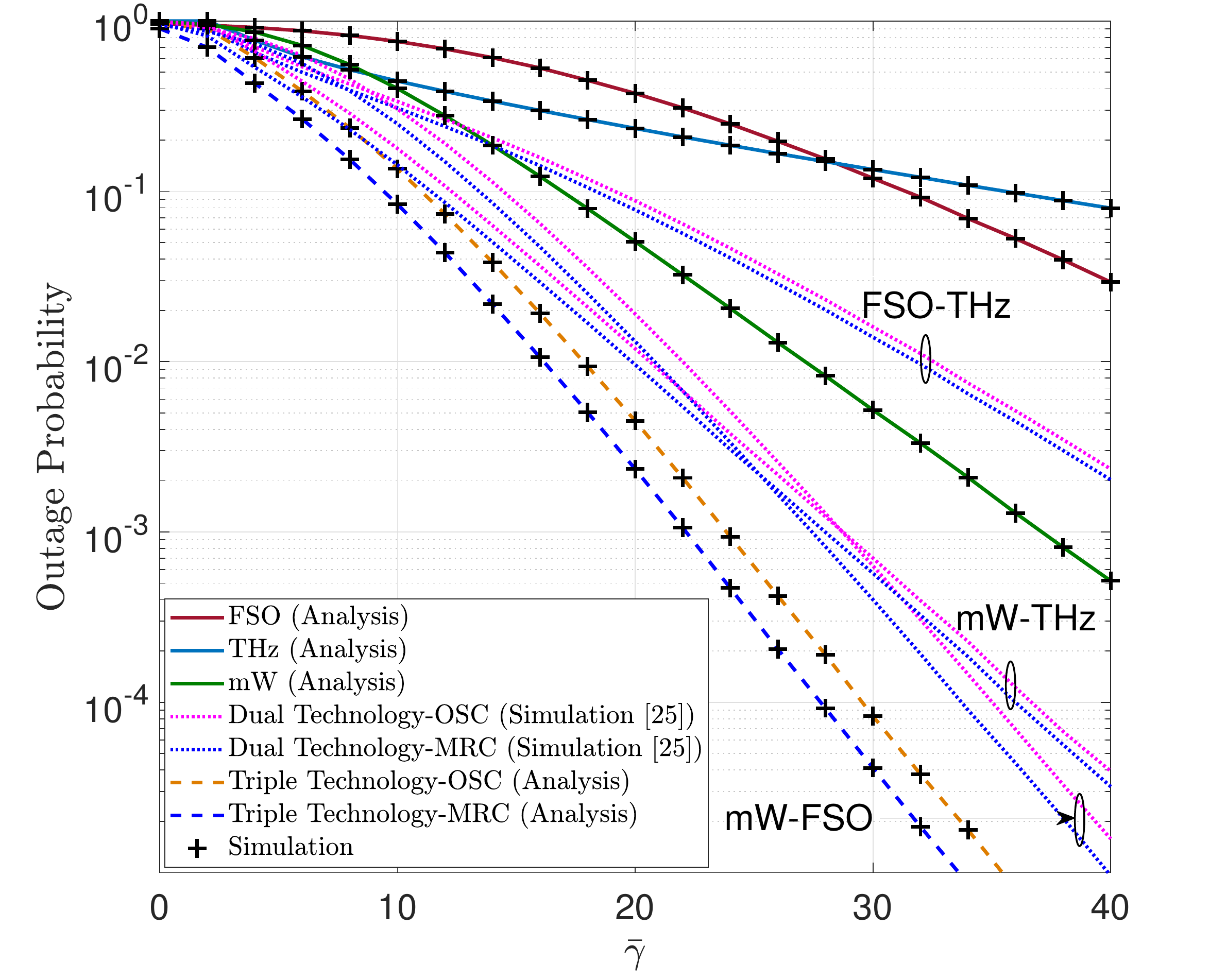}} 
	\subfigure[Outage probability Comparison with existing literature results]{\includegraphics[scale=0.35]{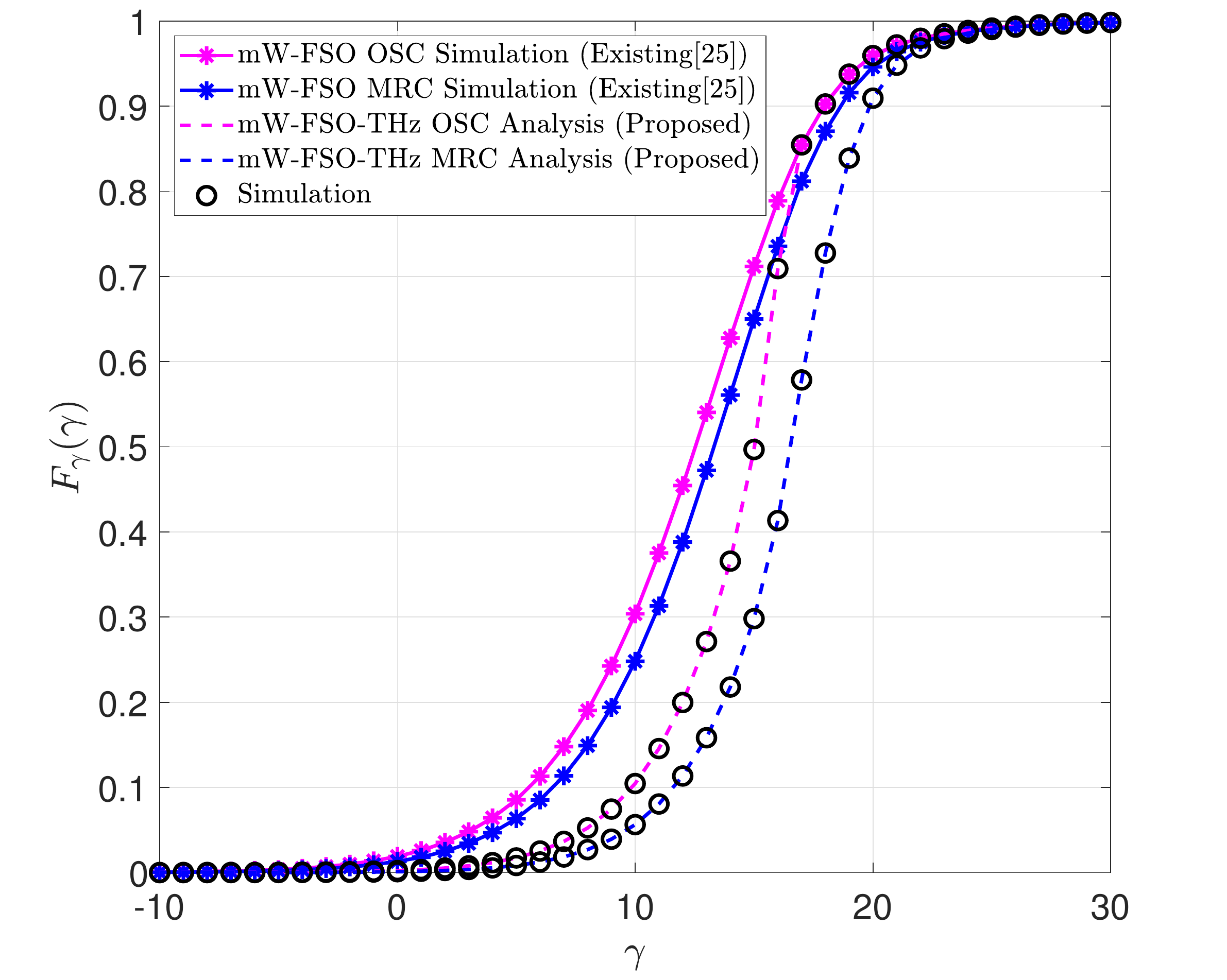}} 
	\caption{Outage probability performance of the Hybrid BHL with $ m=0.5, \Delta=0.9, k=1, \mu_i=0, \sigma_i=0.6$ and strong atmospheric turbulence.}
	\label{fig:outage_hybrid}	
\end{figure*}

\begin{figure}	
	\includegraphics[width=\columnwidth]{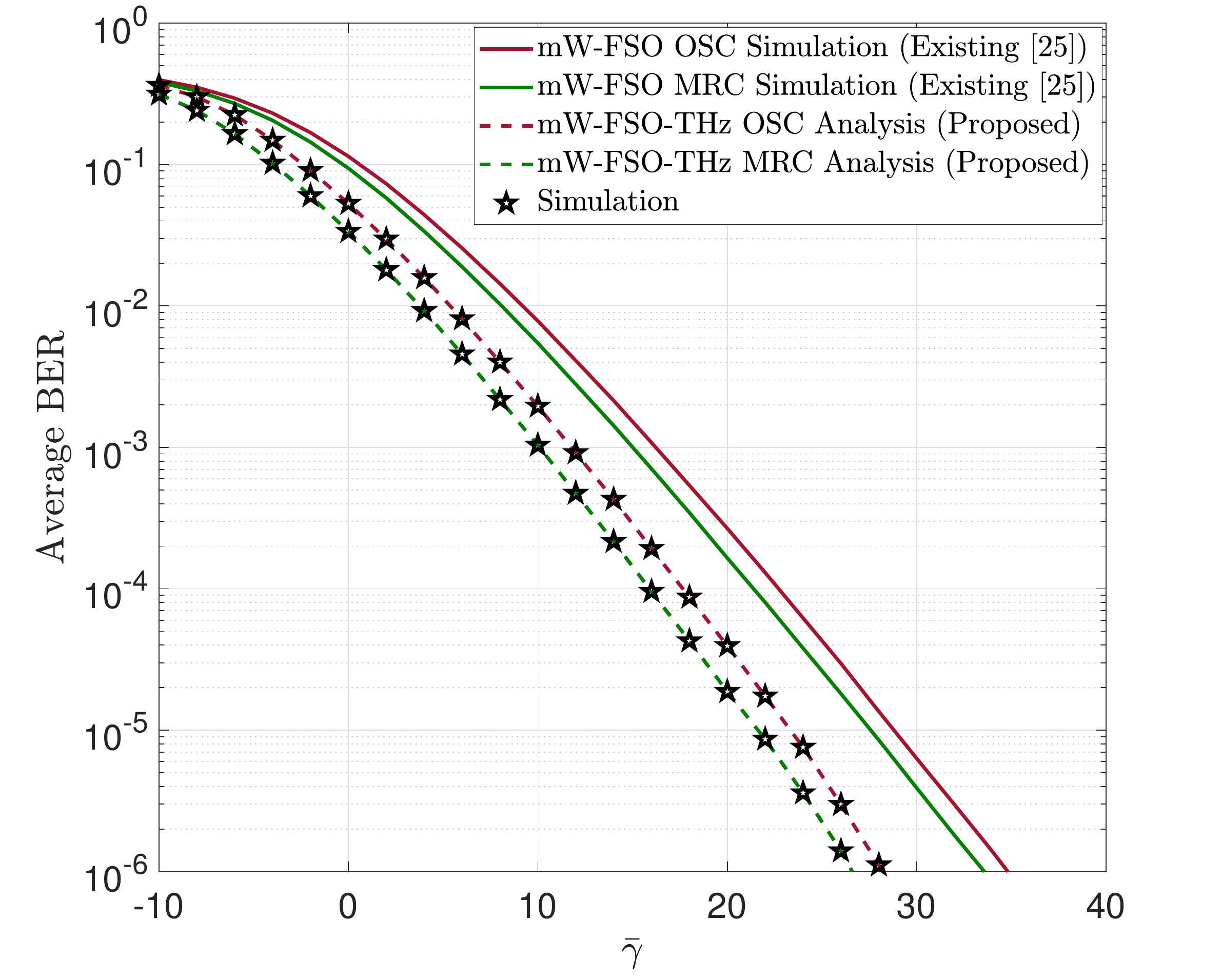}
	\vspace{-0mm}
	\caption{Average BER performance of the hybrid BHL with $m=0.5, \Delta=0.9, k=1, \mu_i=0, \sigma_i=0.8$ and strong atmospheric turbulence.}
	\label{fig:ber_hybrid}
\end{figure}

\begin{figure*}[t]
	\centering
	\subfigure[Outage probability]{\includegraphics[scale=0.35]{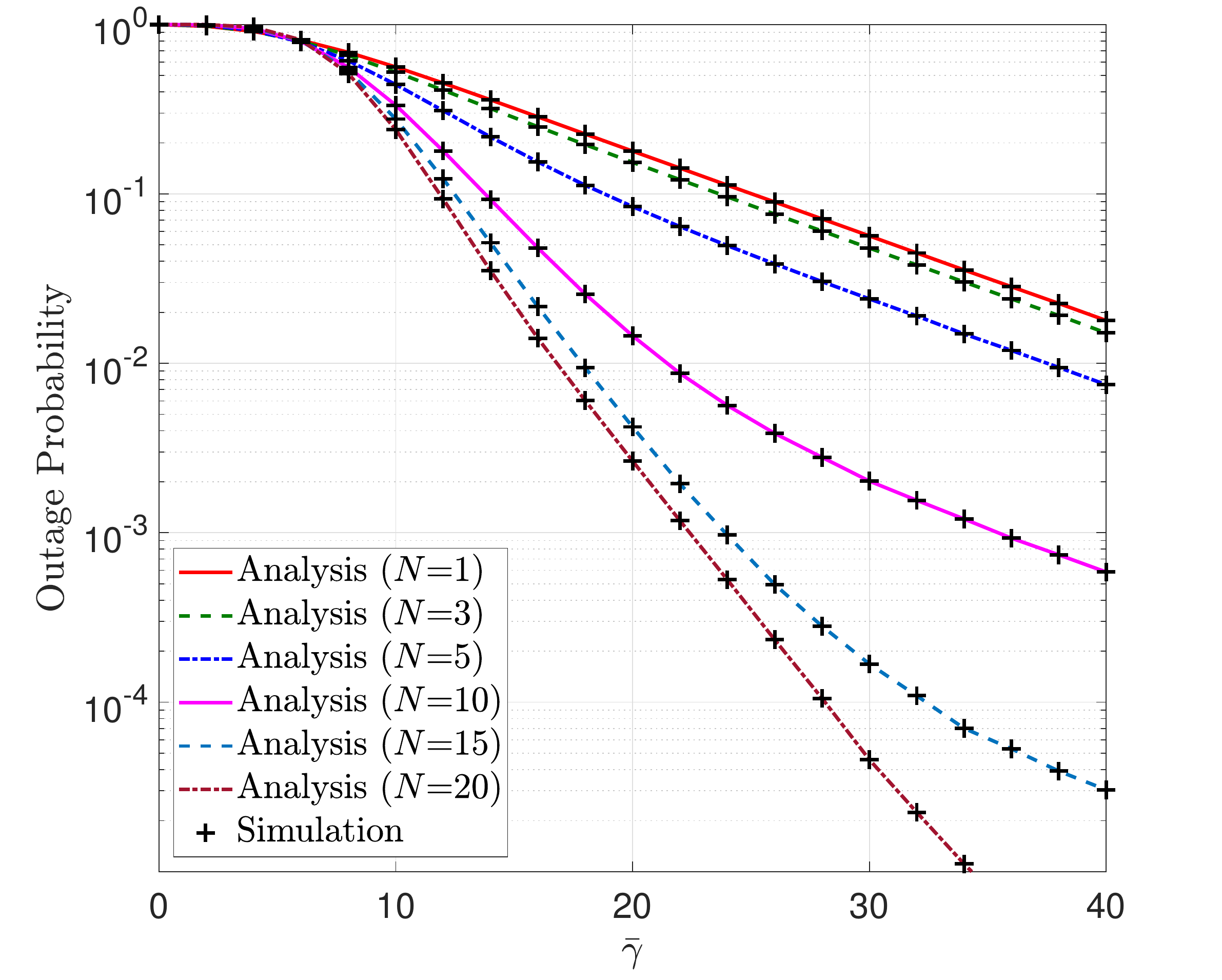}} 
	\subfigure[Average BER]{\includegraphics[scale=0.35]{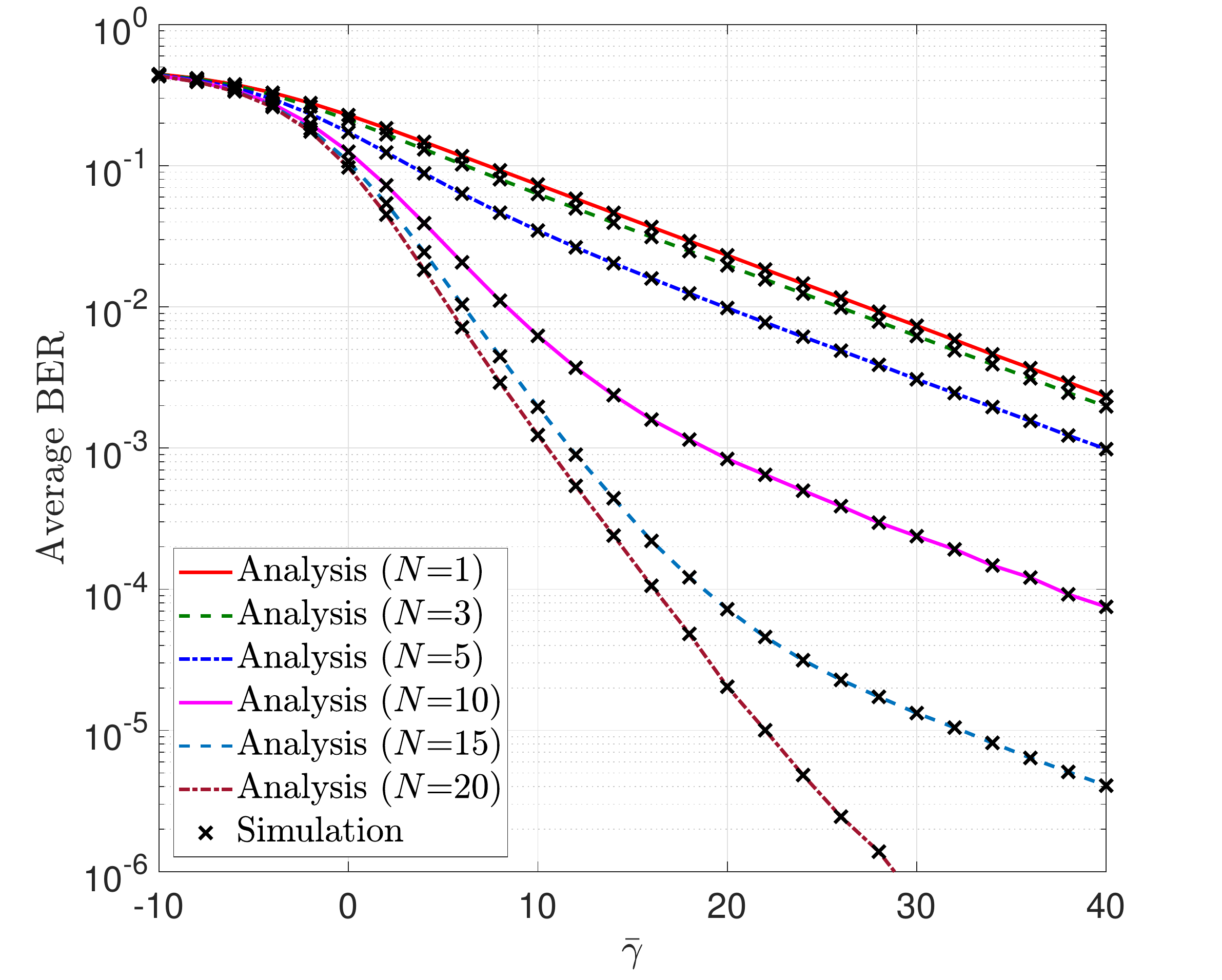}} 
	\caption{Outage probability and average BER performance of integrated link with $m=0.5, \Delta=0.9, k=1, \mu_i=0, \sigma_i=0.6$ and strong turbulence with varying number of IoT devices in the access link.}
	\label{fig:outage_ber_df}	
\end{figure*}

\begin{figure}	
	\includegraphics[width=\columnwidth]{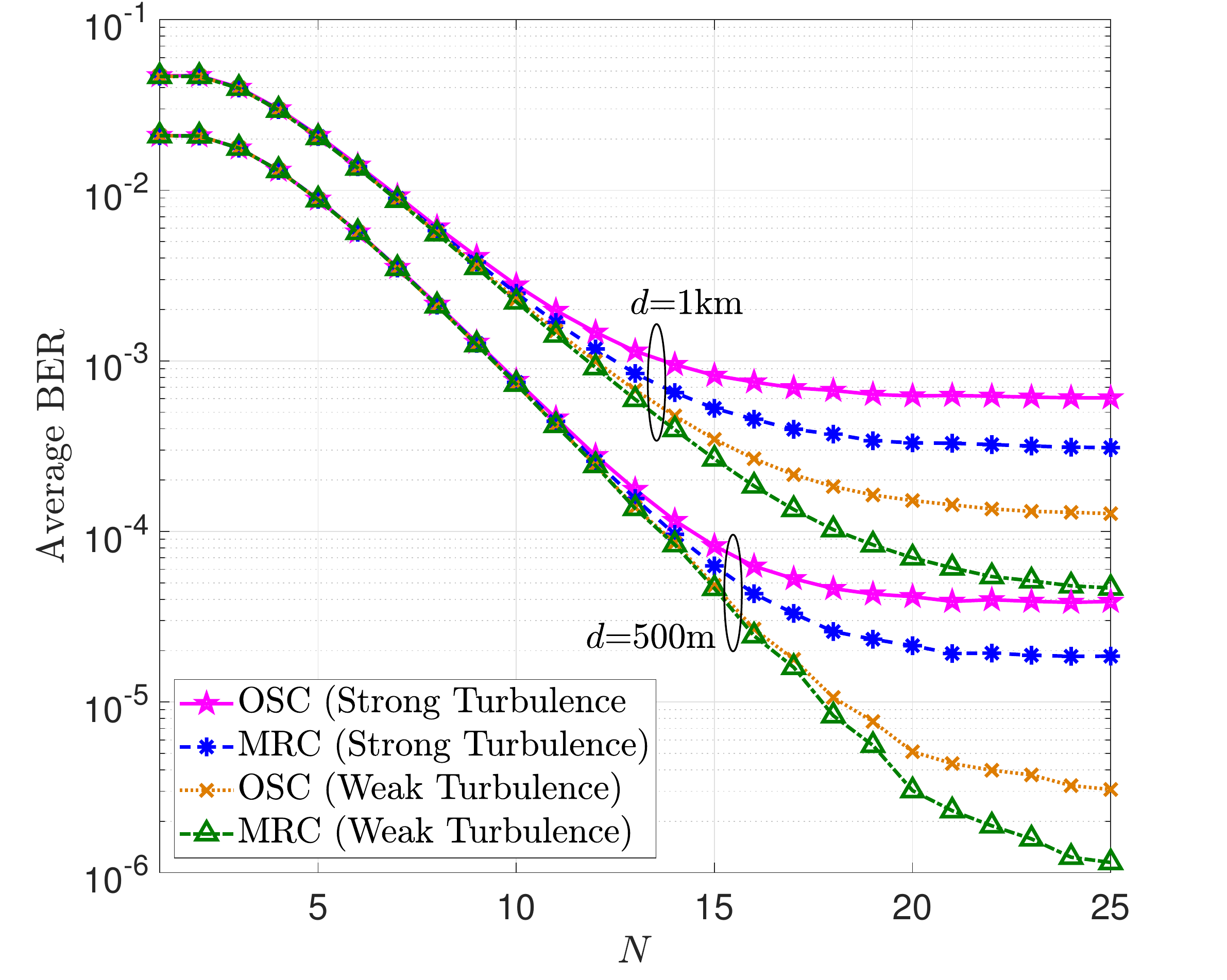}
	\vspace{-0mm}
	\caption{Effect of number of IoT devices on the average BER performance of the integrated link.}
	\label{fig:ber_N}
\end{figure}

\subsection{Average BER}
The average BER refers to the average rate of errors occurring over a given period of time in a communication system and is given by \cite{Ansari2011}
\begin{flalign} \label{eq:ber_general_eqn}
	\bar{P}_e = \frac{q^p}{2\Gamma(p)} \int_{0}^{\infty} e^{-q\gamma} \gamma^{p-1} F_{\gamma} (\gamma) d\gamma
\end{flalign}
The selection of modulation scheme is represented by the values of $p$ and $q$, while $F_{\gamma} (\gamma)$ refers to the CDF of the communication system.
Denote  $\bar{P}_e^{\rm IoT}$ and $\bar{P}_e^{\rm IoT}$ as the  average BER of the individual links: IoT and  BHL (either using MRC or OSC), respectively.

\begin{my_lemma} \label{lemma:ber_relay_eqn}
	The average BER of the integrated link is given by
	\begin{flalign} \label{eq:ber_relay_eqn}
	\bar{P}_e^{\rm IL} = \bar{P}_e^{\rm IoT} + \bar{P}_e^{\rm BHL} - \bar{P}_e^{\rm IoT} \bar{P}_e^{\rm BHL}
\end{flalign}
where
\begin{flalign} \label{eq:ber_mu}
	\bar{P}_e^{\rm IoT} = \frac{q^p}{2}\sum_{k=0}^{N} {N \choose k}  (-1)^k \bigg(\frac{k \lambda \gamma }{{\bar{\gamma}}^{\rm RF}} + q\bigg)^{-p} 
\end{flalign}
and 
\begin{flalign} \label{eq:ber_hybrid_mrc}
	&\bar{P}_e^{\rm BHL} = \frac{m^m }{2\Gamma(m)\Gamma(p)}\sum \limits_{j = 0}^\infty {\frac{K^jd_j q^{-j} }{{(\Gamma (j + 1))^2(2\zeta ^2)^{j+1}} ({{\bar{\gamma}}^{\rm mW}})^{j+1}}} \nonumber \\ & \times \sum_{i=1}^{K} \frac{\rho^2 w_i \exp\Big(\frac{-\mu_i^2}{2 \sigma_i^2}\Big) }{2\sqrt{2\pi}\sigma_i ({\bar{\gamma}}^{\rm THz})^{\frac{1}{2}}}  \frac{\alpha\phi}{\psi\Gamma(\alpha)\Gamma(\beta)} \nonumber \\ &  H^{0,5:1,1;2,2;1,0;1,0}_{5,4:1,1;3,2;0,1;0,1} \left[\begin{matrix} \frac{\gamma \zeta^{-2}}{2q {\bar{\gamma}}^{\rm mW}}, \frac{\alpha  \gamma^{\frac{1}{2}} }{\psi {(q{\bar{\gamma}}^{\rm FSO})}^{\frac{1}{2}} } ,
		\frac{\gamma q^{-1}}{{\bar{\gamma}}^{\rm THz}} , \frac{-2\mu_i \gamma^{\frac{1}{2}}}{(q{\bar{\gamma}}^{\rm THz})^{\frac{1}{2}}} \end{matrix} \Bigg| \begin{matrix} T_7 \\ T_8	\end{matrix}\right]
\end{flalign}
where $T_7 = \bigl\{(\rho-1;0,0,2,1)^2, (\frac{1}{2}; 0,0, 1, \frac{1}{2}), (-1-j;1,\frac{1}{2},1,\frac{1}{2}), (-1-p-j;1,\frac{1}{2},1,\frac{1}{2})\bigr\}: \bigl\{(-j,1)\bigr\} ; \bigl\{(-\beta,1),(\frac{1}{2},\frac{1}{2}) (\phi,1)\bigr\} ;\bigl\{-\bigr\} ; \bigl\{-\bigr\} ;  $ and $T_8 = \bigl\{(\rho-2;0,0,2,1)^2, (-j;1,\frac{1}{2},1,\frac{1}{2}), (-2-j;1,\frac{1}{2},1,\frac{1}{2})\bigr\}: \bigl\{(0,1)\bigr\} ; \bigl\{(\beta-1,1), (\phi,1)\bigr\} ; \bigl\{(0,1)\bigr\};\bigl\{(0,1)\bigr\} $.

\end{my_lemma}

\begin{IEEEproof}
The average BER of DF relaying using Gray coding can be represented as follows: \cite{Tsiftsis2006}:
	\begin{flalign} \label{eq:ber_relay_eqn2}
\bar{P}_e^{\rm IL} = \bar{P}_e^{\rm IoT} + \bar{P}_e^{\rm BHL} - \bar{P}_e^{\rm IoT} \bar{P}_e^{\rm BHL}
\end{flalign}
The derivation of the average BER for individual links are  presented in Appendix C.
\end{IEEEproof}

Likewise, the ergodic capacity of the integrated link can also be derived by following the similar procedure but omitted to avoid redundancy.

\section{Simulation and Numerical Results}

In this section, we demonstrate the performance of the considered system by assessing the performance the hybrid BHL and its integration with the IoT. We verify the accuracy of our derived analytical expressions through Monte Carlo simulations. We consider $50$ terms for the convergence of the infinite series for the FTR channel in the mW band. We assume perfect channel knowledge at the base station to implement the BHL-MRC. To ensure consistency in our comparison, we maintained the same average SNR for all three links in the BHL i.e., $\bar{\gamma}^{\rm mW}= \bar{\gamma}^{\rm FSO}=\bar{\gamma}^{\rm THz}=\bar{\gamma}$. As a result, the system's performance is determined by level of randomness rather than the path-gain of each individual channel. Our evaluation of the IoT network includes a channel coherence time of $1$ \mbox{ms} and a data packet size of $1$ \mbox{kB}  for each device. We will showcase the BHL link's effectiveness in different situations, and then examine the integrated link's performance between the IoT and BHL. The list of simulation parameters is provided in Table \ref{tab:simulation_parameters}.

First, in Fig. \ref{fig:outage_hybrid}, we demonstrate the outage probability performance of the hybrid BHL-MRC and BHL-OSC with specific parameters ($ m=0.5, \Delta=0.9, k=1, \mu_i=0, \sigma=0.6$). In particular, Fig. \ref{fig:outage_hybrid}(a) compares the proposed BHL's performance with individual link and any dual-technology BHL, demonstrating that the proposed BHL outperforms both options. It can be observed from the figure that the system's outage performance improves almost $10$ times for the triple-technology BHL-MRC than the existing dual-technology (mW-FSO) for an average SNR of 30 \mbox{dB}. Fig. \ref{fig:outage_hybrid}(b) compares the outage probability of the proposed BHL schemes with an existing mW-FSO BHL\cite{Badarneh2020_mW_FSO_simultaneous}, highlighting that the dual-technology BHL has a higher outage probability for a given value of average SNR. 
 
Next, Fig. \ref{fig:ber_hybrid} shows the average BER of the proposed BHL schemes compared to the existing mmWave-FSO BHL. The figures demonstrate that the proposed scheme has a better average BER than the existing dual-technology BHL counterpart for a wide the range of average SNR. The plots clearly indicate that employing a triple-technology backhaul instead of an existing dual-technology BHL can result in a saving of $5$ \mbox{dB} of transmitted power to achieve an average BER in the range of $10^{-3}$-$10^{-5}$.

In Fig. \ref{fig:outage_ber_df}, we demonstrate how the number of IoT devices impacts the outage probability and average BER performance of the integrated link. Fig. \ref{fig:outage_ber_df}, displays the effect of  outage probability and average BER performance of the mixed hybrid BHL-MRC and the IoT access link. As shown in Fig. \ref{fig:outage_ber_df}(a), increasing the number of IoT devices  improves the outage performance of the integrated system. At first, increasing the number of devices from $1$ to $5$ does not have a significant impact on the performance, but as the number of devices increases further, the average BER performance improves considerably. This suggests that having more devices can mitigate the randomness of the wireless channel, leading to a more reliable communication link. Similarly, Fig. \ref{fig:outage_ber_df}(b) depicts a similar trend in the average BER performance as the number of IoT devices in the access link increases.

In Fig. \ref{fig:ber_N}, we present the average BER performance of the integrated link using realistic simulation parameters, as specified in Table \ref{tab:simulation_parameters}. We investigate the system performance with different distances for the BHL and varying numbers of devices in the IoT network. Specifically, the distance between the IoT devices and the UAV is set at $100$ \mbox{m}, while two distances are considered for the BHL, i.e., $500$ \mbox{m} and $ 1 $ \mbox{km}. For the mmWave link, we employ the 3GPP path-loss model given by $H_{l_{\rm mW}} = 32.4 + 17.3\log_{10}(d) + 20\log_{10}(10^{-9}f)$. This results in path-loss values of $ 113 $ \mbox{dB} at $500$ \mbox{m} and $ 118.27 $ \mbox{dB} for the mmWave backhaul at a carrier frequency $50$ \mbox{GHz}. The visibility range of the FSO link is adjusted to $ 370 $ \mbox{m} and $ 620 $ \mbox{m}, corresponding to the BHL distances of $500$ \mbox{m} and $ 1 $ \mbox{km}, respectively. We consider both weak and strong turbulence scenarios for the FSO link. The path-loss for the FSO link is determined using $H_{l_{\rm FSO}} = \exp(-\psi d)$, where $\psi$ is the attenuation coefficient \cite{Badarneh2020_mW_FSO_simultaneous}, resulting in path-loss values of $ 45.89 $ \mbox{dB} for $500$ \mbox{m} and $ 52.78 $ \mbox{dB} for $ 1 $ \mbox{km}. For the THz link, the path loss due to the atmospheric absorption  $H_{l_{\rm THz}} = \exp\left(-\frac{1}{2}\kappa d\right)$, where $\kappa$ is the absorption coefficient  \cite{Boulogeorgos_Analytical},  yields path-loss values of $ 136.65 $ \mbox{dB} for $ 500 $ \mbox{m} and $ 143.36 $ \mbox{dB} for $ 1 $ \mbox{km}. For the AN to the UAV  link,  the  3GPP path-loss model results into a path-loss value of $ 65 $ \mbox{dB} for an average IoT link distance of $ 100 $ \mbox{m}. 	We fix the transmit power at $ 0 $ \mbox{dBm}, and utilizing the parameters provided in Table \ref{tab:simulation_parameters}, we obtain an average SNR of $ 20 $ \mbox{dB} ($ 500 $ \mbox{m}) and 13 \mbox{dB} ($ 1 $ \mbox{km}) for all three links in the backhaul.  To maintain the average SNR of the RF link at the same level, the transmitted power of the IoT devices should be reduced, resulting in a power saving by $ 23 $ \mbox{dBm} for $500$ \mbox{m} and $ 30 $ \mbox{dBm}  for $1$\mbox{km} distance. Note that we ignore the local communication between IoT devices to the AN.

The plots in Fig.~5 demonstrate that the average BER performance of the integrated link improves as the number of devices in the IoT network increases. However, after approximately 20 devices, the average BER reaches a saturation point (which is a characteristic of the OSC technique).  It is evident from the plots that average BER increases as we increase the distance of the BHL owing to the greater atmospheric attenuation and path-loss in the BHL.  Further, we can illustrate the achievable data rate using the simulation scenario of Fig.~5. In Fig.~5,  an average SNR for $20$ \mbox{dB} gives an spectral efficiency of $8.28$ \mbox{bits/sec/Hz} for the MRC and $7.71$ \mbox{bits/sec/Hz} for the OSC. Similarly, setting parameters for the IoT network such that an average SNR the AN is $20$ \mbox{dB} (same average SNR is required for dual-hop DF system to have more meaningful results of the relay-assisted system), giving an achievable spectral efficiency of  $6.65$ \mbox{bits/sec/Hz}. Thus, the backhaul link is sufficient to support the data collection from $N$=$20$ devices in the IoT network, each device having a packet size of $332$ \mbox{kb} of data if $1$ MHz of channel bandwidth is assumed for the AN.

The hybrid backhaul based data collection network has the capability of delivering superior performance, but in exchange for this benefit, there exists a trade-off between performance and complexity which varies across different operating conditions. Therefore, we have conducted comprehensive simulations to capture the behavior of the integrated link, and presented a comparison between the proposed wireless BHL and other existing technologies in Table \ref{table:comparison} on page \pageref{table:comparison}. We have created various use-cases to identify the most suitable choice between single, dual, and triple-technology BHL for different listed scenarios. We consider two types of channel conditions, namely Type-1 (poor) and Type-2 (good), for each of the mW (Type-1: $m=0.5, \Delta=0.9$; Type-2: $m=20, \Delta=0.1$), FSO (Type-1: strong turbulence; Type-2: weak turbulence), and THz (Type-1: $\sigma_i = 0.6$; Type-2: $\sigma_i = 1.1$) links, individually, as well as and in all possible combinations for low, moderate, and high SNR regions. The last column of the each row shows the outage probability of the integrated link with the triple-technology BHL-MRC, while the other columns in that row indicate the outage probability of the particular scenario as the factor of the triple-technology BHL-MRC. This allows us to identify the best alternative to the triple-technology BHL-MRC based on the lowest factor in a single row for a given SNR range. The minimum factor for each row, for both single-technology and dual-technology BHL-MRC combinations, is highlighted. For instance, if we consider Type-1 channel for all the individual links, the mW link alone provides a better outage probability at moderate and high SNR, while the THz link performs better at low SNR for single-technology BHL. The mW-THz BHL outperforms the mW-FSO in low and moderate SNR, while the mW-FSO performs better in high SNR for the dual-technology BHL. However, the decision to choose a BHL technology depends on the specific network configuration and the acceptable level of outage probability, with a trade-off between network complexity and performance. Choosing a single, dual, or triple-technology BHL based on the desired outage probability can reduce network complexity and power requirements. Nevertheless, the triple-technology BHL consistently delivers superior performance regardless of the channel and SNR scenarios.

\begin{table*}[tp] 	

	\caption{Comparison of outage probability for integrated link using triple, dual, and single-technology based backhaul with $N=5$ devices in the IoT network.}
		\small{The Table can be read as follows: the outage probability of the triple-technology backhaul (given in the last column) is denoted by $P_{\rm out}$, while $X$ is the factor by which the outage probability of a single or dual-technology backhaul exceeds that of the triple-technology backhaul. Thus, to get the outage probability of different configurations $P_{\rm out}$ should be multiplied by the corresponding factor $X$, depicted in rows. For example, $P_{\rm out}$ (for mW at 5 \mbox{dB} SNR with Type-1 channel) is $2.34 \times 0.34=0.79$.}
	\label{table:comparison}	
	\begin{center}
		\begin{tabular}{|p{2.1cm}||p {1.4cm}|p {1.4cm}|p {1.4cm}||p {1.5cm}|p {1.5cm}|p {1.7cm}||p {2.3cm} p {0.1mm}|} 
			\hline   
			\backslashbox{\textbf{SNR}}{\textbf{Link}}  & \centering\textbf{mW} \centering\ $X \times P_{ \rm out} $ & \centering\textbf{FSO}\centering\ $X \times P_{ \rm out} $ &  \centering\textbf{THz} \centering\ $X \times P_{ \rm out} $& \textbf{mW-FSO} \centering\ $X \times {\rm P_{ \rm out}} $ &  \textbf{mW-THz} \centering\ $X \times P_{ \rm out} $ & \textbf{FSO-THz}\centering\ $X \times P_{ \rm out} $ &\textbf{mW-FSO-THz}  \centering $ P_{ \rm out}$&  \\  
			\hline  
			\centering Low  ($ 5 \mbox{dB} $) &  2.34 (\centering${\rm Type}$-$\rm{1}$)  & 2.64 (\centering${\rm Type}$-$\rm{1}$)&  \cellcolor{red!25} 2.00 (\centering${\rm Type}$-$\rm{1}$) & \centering1.91 & \centering\cellcolor{blue!25} 1.27 & \centering{ 1.67 } & \cellcolor{green!25} \centering {0.34} &  \cellcolor{green!25} \\  
			\hline
			\centering Low  ($ 5 \mbox{dB} $) & \cellcolor{red!25} 2.94 (\centering${\rm Type}$-$\rm{1}$) & 3.10 (\centering${\rm Type}$-$\rm{1}$) & 3.15 (\centering${\rm Type}$-$\rm{2}$) & \centering2.41 & \centering\cellcolor{blue!25}1.42  & \centering 2.30 & \cellcolor{green!25} \centering {0.27}  &  \cellcolor{green!25} \\  
			\hline
			\centering Low  ($ 5 \mbox{dB} $) &  2.92 (\centering${\rm Type}$-$\rm{1}$)  & 3.10 (\centering${\rm Type}$-$\rm{2}$) & \cellcolor{red!25} 2.49 (\centering${\rm Type}$-$\rm{1}$) & \centering2.02 & \centering\cellcolor{blue!25}1.58  & \centering 1.81 & \cellcolor{green!25} \centering {0.27}  &  \cellcolor{green!25} \\  
			\hline
			\centering Low  ($ 5 \mbox{dB} $) & \cellcolor{red!25} 4.42 (\centering${\rm Type}$-$\rm{1}$) & 4.67 (\centering${\rm Type}$-$\rm{2}$) & 4.73 (\centering${\rm Type}$-$\rm{2}$) & \centering3.1 & \centering\cellcolor{blue!25}2.12 & \centering 2.63  & \cellcolor{green!25} \centering {0.18} &  \cellcolor{green!25} \\  
			\hline
			\centering Low  ($ 5 \mbox{dB} $) &  2.42 (\centering${\rm Type}$-$\rm{2}$)  & 2.73 (\centering${\rm Type}$-$\rm{1}$) & \cellcolor{red!25} 2.10 (\centering${\rm Type}$-$\rm{1}$) & \centering1.96 &\centering \cellcolor{blue!25}1.3  & \centering 1.72  & \cellcolor{green!25} \centering {0.33} &  \cellcolor{green!25} \\  
			\hline
			\centering Low  ($ 5 \mbox{dB} $) & \cellcolor{red!25} 3.19 (\centering${\rm Type}$-$\rm{2}$) & 3.59 (\centering${\rm Type}$-$\rm{1}$) & 3.40 (\centering${\rm Type}$-$\rm{2}$) & \centering2.58 & \centering\cellcolor{blue!25}1.43 & \centering 2.48 & \cellcolor{green!25} \centering {0.25} &  \cellcolor{green!25} \\  
			\hline
			\centering Low  ($ 5 \mbox{dB} $) &  3.00 (\centering${\rm Type}$-$\rm{2}$) & 3.23 (\centering${\rm Type}$-$\rm{2}$) & \cellcolor{red!25} 2.62 (\centering${\rm Type}$-$\rm{1}$) & \centering2.21 & \centering\cellcolor{blue!25}1.65  & \centering 1.89 & \cellcolor{green!25} \centering {0.26} &  \cellcolor{green!25} \\  
			\hline
			\centering Low  ($ 5 \mbox{dB} $) & \cellcolor{red!25} 4.40 (\centering${\rm Type}$-$\rm{2}$) & 4.62 (\centering${\rm Type}$-$\rm{2}$) & 4.68 (\centering${\rm Type}$-$\rm{2}$) & \centering3.15 & \centering\cellcolor{blue!25}1.96  & \centering 2.60  & \cellcolor{green!25} \centering {0.18} &  \cellcolor{green!25} \\  
			\hline
			\hline
			\centering Moderate  ($15\mbox{dB} $) & \cellcolor{red!25} 9.94 (\centering${\rm Type}$-$\rm{1}$) & 35.31 (\centering${\rm Type}$-$\rm{1}$) & 20.00 (\centering${\rm Type}$-$\rm{1}$) & \centering4.00 & \centering\cellcolor{blue!25}2.48  & \centering 10.06 & \cellcolor{green!25} \centering {0.016} &  \cellcolor{green!25} \\
			\hline
			\centering Moderate  ($15\mbox{dB} $) &  53.00 (\centering${\rm Type}$-$\rm{1}$) & 188.33 (\centering${\rm Type}$-$\rm{1}$) & \cellcolor{red!25} 31.00 (\centering${\rm Type}$-$\rm{2}$) & \centering21.33 & \centering\cellcolor{blue!25}2.77  & \centering 13.77 & \cellcolor{green!25} \centering {0.003} &  \cellcolor{green!25} \\
			\hline
			\centering Moderate  ($15\mbox{dB} $) & \cellcolor{red!25} 31.80 (\centering${\rm Type}$-$\rm{1}$) & 54.00 (\centering${\rm Type}$-$\rm{2}$) & \centering 64.00 (\centering${\rm Type}$-$\rm{1}$) & \centering\cellcolor{blue!25}4.49 & \centering7.8  & \centering 13.82 & \cellcolor{green!25} \centering {0.005} &  \cellcolor{green!25} \\
			\hline
			\centering Moderate  ($15\mbox{dB} $) &  169.15 (\centering${\rm Type}$-$\rm{1}$) & 287.23 (\centering${\rm Type}$-$\rm{2}$) & \cellcolor{red!25} 98.94 (\centering${\rm Type}$-$\rm{2}$)& \centering23.89 & \centering\cellcolor{blue!25}8.83  & \centering 16.28 & \cellcolor{green!25} \centering {9.4$\times$$10^{-4}$} &  \cellcolor{green!25} \\
			\hline
			\centering Moderate  ($15\mbox{dB} $) & \cellcolor{red!25} 10.00 (\centering${\rm Type}$-$\rm{2}$) & 48.46 (\centering${\rm Type}$-$\rm{1}$) & 27.45 (\centering${\rm Type}$-$\rm{1}$) & \centering4.18 & \centering\cellcolor{blue!25}2.56  & \centering 13.81 & \cellcolor{green!25} \centering {11.66$\times$$10^{-3}$} &  \cellcolor{green!25} \\
			\hline
			\centering Moderate  ($15\mbox{dB} $) &  50.87 (\centering${\rm Type}$-$\rm{2}$) & 245.65 (\centering${\rm Type}$-$\rm{1}$) & \cellcolor{red!25} 40.43 (\centering${\rm Type}$-$\rm{2}$) & \centering21.17 & \centering\cellcolor{blue!25}2.78  & \centering 17.96 & \cellcolor{green!25} \centering {2.3$\times$$10^{-3}$} &  \cellcolor{green!25} \\
			\hline
			\centering Moderate  ($15\mbox{dB} $) & \cellcolor{red!25} 31.62 (\centering${\rm Type}$-$\rm{2}$) & 72.97 (\centering${\rm Type}$-$\rm{2}$) & 86.49 (\centering${\rm Type}$-$\rm{1}$) & \centering\cellcolor{blue!25}4.54 & \centering8.05  & \centering 18.68 & \cellcolor{green!25} \centering {3.7$\times$$10^{-3}$} &  \cellcolor{green!25} \\
			\hline
			\centering Moderate  ($15\mbox{dB} $) & 169.57 (\centering${\rm Type}$-$\rm{2}$) & 391.30 (\centering${\rm Type}$-$\rm{2}$) & \cellcolor{red!25} 134.78 (\centering${\rm Type}$-$\rm{2}$) & \centering24.35 & \centering\cellcolor{blue!25}9.28   & \centering 22.17 & \cellcolor{green!25} \centering {6.9$\times$$10^{-4}$} &  \cellcolor{green!25} \\
			\hline
			\hline
			\centering High  ($30\mbox{dB} $)  & \cellcolor{red!25} 139.47 (\centering${\rm Type}$-$\rm{1}$) & 3157.90 (\centering${\rm Type}$-$\rm{1}$) & 3526.30 (\centering${\rm Type}$-$\rm{1}$) & \centering\cellcolor{blue!25}11.13 & \centering14.74  & \centering368.42  & \cellcolor{green!25} \centering {3.8$\times$$10^{-5}$} &  \cellcolor{green!25} \\
			\hline
			\centering High  ($30\mbox{dB} $)  & \cellcolor{red!25} 5300 (\centering${\rm Type}$-$\rm{1}$) & 120000 (\centering${\rm Type}$-$\rm{1}$) & 8000 (\centering${\rm Type}$-$\rm{2}$) & \centering423  & \centering\cellcolor{blue!25}23.00 & \centering 700 & \cellcolor{green!25} \centering {1$\times$$10^{-6}$} &  \cellcolor{green!25} \\
			\hline
			\centering High  ($30\mbox{dB} $)  &  \cellcolor{red!25} 2650 (\centering${\rm Type}$-$\rm{1}$) & 4080 (\centering${\rm Type}$-$\rm{2}$) & 67000 (\centering${\rm Type}$-$\rm{1}$) & \centering\cellcolor{blue!25}8.00 & \centering280 & \centering 421 & \cellcolor{green!25} \centering {2$\times$$10^{-6}$} &  \cellcolor{green!25} \\
			\hline
			\centering High  ($30\mbox{dB} $)  &  \cellcolor{red!25} 26500 (\centering${\rm Type}$-$\rm{1}$) & 40800 (\centering${\rm Type}$-$\rm{2}$) & 40000 (\centering${\rm Type}$-$\rm{2}$) & \centering\cellcolor{blue!25}80 & \centering 115  & \centering235 & \cellcolor{green!25} \centering {2$\times$$10^{-7}$} &  \cellcolor{green!25} \\
			\hline
			\centering High  ($30\mbox{dB} $)  &  \cellcolor{red!25} 119.35 (\centering${\rm Type}$-$\rm{2}$) & 3871 (\centering${\rm Type}$-$\rm{1}$) & 4322.6 (\centering${\rm Type}$-$\rm{1}$) & \centering\cellcolor{blue!25}8.94 & \centering13.23   & \centering 451.61  & \cellcolor{green!25} \centering {3.1$\times$$10^{-5}$} &  \cellcolor{green!25} \\
			\hline
			\centering High  ($30\mbox{dB} $)  &  \cellcolor{red!25} 9250 (\centering${\rm Type}$-$\rm{2}$) & 300000 (\centering${\rm Type}$-$\rm{1}$) & 20000 (\centering${\rm Type}$-$\rm{2}$) & \centering692.50 & \centering\cellcolor{blue!25}40.00   & \centering1750  & \cellcolor{green!25} \centering {4$\times$$10^{-7}$} &  \cellcolor{green!25} \\
			\hline
			\centering High  ($30\mbox{dB} $)  &  \cellcolor{red!25} 1850 (\centering${\rm Type}$-$\rm{2}$) & 4080 (\centering${\rm Type}$-$\rm{2}$) & 67000 (\centering${\rm Type}$-$\rm{1}$) & \centering\cellcolor{blue!25}5.00 & \centering205  & \centering 421 & \cellcolor{green!25} \centering {2$\times$$10^{-6}$} &  \cellcolor{green!25} \\
			\hline
			\centering High  ($30\mbox{dB} $)  &  37000 (\centering${\rm Type}$-$\rm{2}$) & 81600 (\centering${\rm Type}$-$\rm{2}$) &  \cellcolor{red!25} 8000 (\centering${\rm Type}$-$\rm{2}$) & \centering\cellcolor{blue!25}100 & \centering160  & \centering 470 & \cellcolor{green!25} \centering {1$\times$$10^{-7}$} &  \cellcolor{green!25} \\
			\hline
		\end{tabular}

	\end{center}
	
\end{table*}

\section{Conclusions}
We analyzed the end-to-end performance of data collection from an IoT network located in hard-to-reach areas, which is transferred to the core network through an intermediate UAV and a high-speed wireless BHL. We designed a self-configuring algorithm for AN selection from IoT devices in the network. Additionally, we utilized a novel hybrid transmission technique for wireless backhaul, which transmitted data packets simultaneously on mW, FSO, and THz technologies using OSC and MRC. We assessed the outage probability and average BER performance of the integrated link that comprises the IoT and BHL. Moreover, we derived a simpler asymptotic expression for the outage probability in the high SNR region and determined the diversity order for the integrated link.

Our study provides a comprehensive understanding of the selection of hybrid schemes in different channel and SNR scenarios, based on the trade-off between acceptable outage probability and the complexity of the hybrid backhaul. In particular, we found that the mW-THz BHL outperforms the mW-FSO in low and moderate SNR scenarios, while the mW-FSO performs better in high SNR for the dual-technology BHL. The triple-technology BHL consistently delivers superior performance regardless of the channel and SNR scenarios. By leveraging the collective strength of a triple-technology BHL and multiple IoT devices, it may be possible to enhance the overall performance and reliability of the communication link for data collection. This could have a wide range of potential applications, such as in environmental monitoring, disaster response, military communication, and other scenarios where reliable communication is essential. Our proposed solution and statistical analysis provide valuable insights into the performance of IoT networks and can be useful for the future design and optimization of similar networks. Additionally, it has the potential to steer the development of more practical and robust solutions, addressing beam alignment issues and the increased computational demands at the UAV in real-world situations.

\section*{Appendix A}
Substituting the limits of \eqref{eq:pdf_channel_thz_gm} and \eqref{eq:pdf_pointing_thz_gm}, the joint PDF of $|h_{fp}|= |h_{f}|h_{p}$ can be expressed as \cite{papoulis_2002}

\begin{flalign} \label{eq:pdf_thz_der1}
	f_{h_{fp}}(z) = \int _{0}^{1} \frac {1}{y} f_{h_{p}}(y)  f_{h_{f}}\left ({\frac {z}{y}}\right)  \mathrm {d}y.
\end{flalign}

substituting the PDFs from \eqref{eq:pdf_channel_thz_gm} and \eqref{eq:pdf_pointing_thz_gm} in (\ref{eq:pdf_thz_der1}) and simplifying, we get
\begin{flalign} 
	f_{h_{fp}}(z) =& - \sum_{i=1}^{K} \frac{\rho^2 w_i \exp\Big(\frac{-\mu_i^2}{2 \sigma_i^2}\Big)}{\sqrt{2\pi}\sigma_i}\int_{0}^{1}  y^{\rho-2} \ln(y) \nonumber \\  \times & \exp\Big(\frac{(-\frac{z}{y})^2}{2 \sigma_i^2}\Big) \exp\Big(\frac{(\frac{2z\mu_i}{y})}{2 \sigma_i^2}\Big)  \mathrm {d}y.
\end{flalign}
The closed from solution of the above integral with powers on the exponential is not possible. Thus, using the Meijer's G representation of both the exponential functions, we can re write the equation as
\begin{flalign} \label{eq:pdf_thz_combined_der2}
	f_{h_{fp}}(z) = &  - \sum_{i=1}^{K} \frac{\rho^2 w_i \exp\Big(\hspace{-0.5mm}\frac{-\mu_i^2}{2 \sigma_i^2}\Big)}{\sqrt{2\pi}\sigma_i} \bigg(\hspace{-0.5mm}\frac{1}{2\pi \J}\hspace{-0.5mm}\bigg)^{\hspace{-1mm}2} \int_{\mathcal{L}_1} \hspace{-1mm}\int_{\mathcal{L}_2} \hspace{-1mm} \Gamma(0-s_1)  \nonumber \\  \times &   ({z^2})^{s_1}   \Gamma(0-s_2) ({-2z\mu_i})^{s_2} ds_1 ds_2  \times I_1
\end{flalign}

The inner integral can be solved to
\begin{flalign} \label{eq:pdf_thz_combined_inner}
	I_1 \hspace{-0.5mm} = \hspace{-1mm} \int_{0}^{1} \hspace{-2mm} y^{\rho-2-2s_1-s_2} \ln(y)   \mathrm {d}y =  - \frac{(\Gamma(-\rho+2+2s_1\hspace{-0.5mm}+\hspace{-0.5mm}s_2))^2}{(\Gamma(-\rho+3+2s_1\hspace{-0.5mm}+\hspace{-0.5mm}s_2))^2}
\end{flalign}

Substituting back \eqref{eq:pdf_thz_combined_inner} in \eqref{eq:pdf_thz_combined_der2}, and using the definition of bivariate Fox's H-function \cite{Mittal_1972}, we can write the PDF of THz outdoor channel combined with with THz pointing error in \eqref{eq:pdf_thz_combined}. The CDF can be derived using the PDF in $F_{h_{fp}}^{\rm THz}(z) = \int_{0}^{z} f_{h_{fp}}^{\rm THz}(z) dz$. Using \eqref{eq:pdf_thz_combined}, we can write CDF as
\begin{flalign} \label{eq:cdf_thz_combined_der1}
	&F_{h_{fp}}(z) = \sum_{i=1}^{K} \frac{\rho^2 w_i \exp\Big(\frac{-\mu_i^2}{2 \sigma_i^2}\Big)}{\sqrt{2\pi}\sigma_i} \bigg(\frac{1}{2\pi \J}\bigg)^2 \hspace{-0.5mm} \int_{\mathcal{L}_1} \hspace{-0.5mm} \int_{\mathcal{L}_2} \hspace{-1mm} \Gamma(0-s_1)  \nonumber \\  & \times    ({z^2})^{s_1}   \Gamma(0-s_2) ({-2z\mu_i})^{s_2} \frac{(\Gamma(-\rho+2+2s_1+s_2))^2}{(\Gamma(-\rho+3+2s_1+s_2))^2}  \nonumber \\  & \times ds_1 ds_2 \times I_2
\end{flalign}

The inner integral is straightforward and is simplified to
\begin{flalign} \label{eq:inner_i2}
	I_2 = \int_{0}^{z} z^{2s_1+s_2} dz = \frac{z^{2s_1+s_2+1} \Gamma(2s_1+s_2+1)}{\Gamma(2s_1+s_2+2)}
\end{flalign}

Substituting $I_2$ in \eqref{eq:cdf_thz_combined_der1}, and applying the definition of bivariate Fox's H-function, we get the CDF in \eqref{eq:cdf_thz_combined}. Using the standard transforms of the random variables for the the SNR \cite{papoulis_2002}, we get the PDF and CDF of the SNR to conclude the proof of Theorem 1.

\section*{Appendix B}
To begin with the proof, we compute the MGF for each of the three technologies used in the BHL. To achieve this, we substitute the PDF of \eqref{eq:PDF_F_pe} into \eqref{eq:mgf_eqn}, employ the Mellin Barnes type representation of the exponential function, and change the order of integration to obtain the MGF of the mW link as
\begin{flalign} 
	&\mathcal{M}_{\gamma}^{\rm mW} (s) = \frac{{{m^m}}}{{\Gamma (m)}}\sum \limits_{j = 0}^\infty {\frac{{K^jd_j }}{{(\Gamma (j + 1))^2(2\zeta ^2)^{j+1}} ({{\bar{\gamma}}^{\rm mW}})^{j+1}}}  \nonumber \\ & \times  \frac{1}{2\pi \J} \int_{{L}_3} \Gamma(0-s_3) \Big(\frac{1}{2\zeta ^2 {\bar{\gamma}}^{\rm mW}}\Big)^{s_3} ds_3 \int_{0}^{\infty} e^{-s\gamma} \gamma^{j+s_3} d\gamma
\end{flalign}
which, using the identity \cite[3.381/4]{Gradshteyn} can be simplified to
\begin{flalign} \label{eq:mgf_mW}
	&\mathcal{M}_{\gamma}^{\rm mW} (s) = \frac{{{m^m}}}{{\Gamma (m)}}\sum \limits_{j = 0}^\infty {\frac{{K^jd_j }}{{(\Gamma (j + 1))^2(2\zeta ^2)^{j+1}} ({{\bar{\gamma}}^{\rm mW}})^{j+1}}}  \nonumber \\ & \times \hspace{-0.5mm} \frac{1}{2\pi \J} \hspace{-1mm} \int_{{L}_3} \hspace{-2mm} \Gamma(0-s_3) \Big(\frac{1}{2\zeta ^2 {\bar{\gamma}}^{\rm mW}}\Big)^{\hspace{-0.5mm}s_3}  s^{-1-j-s_3} \Gamma(1\hspace{-0.5mm}+\hspace{-0.5mm}j \hspace{-0.5mm}+\hspace{-0.5mm}s_3)  ds_3
\end{flalign}

Similarly, the MGFs of the FSO and THZ links can be obtained as
\begin{flalign} \label{eq:mgf_fso}
	&\mathcal{M}_{\gamma}^{\rm FSO} (s) = \frac{\alpha_{{\scriptscriptstyle  F}}\phi}{\psi\Gamma(\alpha)\Gamma(\beta)} \nonumber \\ & \times \frac{1}{2\pi \J} \int_{{L}_4} \frac{\Gamma(\beta_{\scriptscriptstyle F}-1-s_4)\Gamma(\phi-1-s_4)\Gamma(1+\beta_{\scriptscriptstyle F}+s_4)}{\Gamma(\phi-s_4)} \nonumber \\ & \times \bigg(\frac{\alpha }{\psi ({\bar{\gamma}}^{\rm FSO})^{\frac{1}{2}}}\bigg)^{s_4} ds_4 \times  s^{-\frac{1}{2}-\frac{ s_4}{2}} \Gamma\bigg(\frac{1}{2}+\frac{ s_4}{2}\bigg)
\end{flalign}
\begin{flalign} \label{eq:mgf_thz}
	&\mathcal{M}_{\gamma}^{\rm THz} (s) = \sum_{i=1}^{K} \frac{\rho^2 w_i \exp\Big(\frac{-\mu_i^2}{2 \sigma_i^2}\Big) }{2\sqrt{2\pi}\sigma_i ({\bar{\gamma}}^{\rm THz})^{\frac{1}{2}}} \bigg(\frac{1}{2\pi \J}\bigg)^{\hspace{-1mm}2}  \hspace{-1mm}\int_{\mathcal{L}_5} \hspace{-1mm}\int_{\mathcal{L}_6} \hspace{-1mm}\Gamma(0-s_5)  \nonumber \\  & \times  \hspace{-1mm}  \bigg(\hspace{-0.5mm}\frac{1}{{\bar{\gamma}}^{\rm THz}}\hspace{-0.5mm}\bigg)^{\hspace{-1mm}s_5}  \hspace{-0.5mm} \Gamma(0\hspace{-0.5mm}-\hspace{-0.5mm}s_6) \bigg(\frac{-2\mu_i}{({\bar{\gamma}}^{\rm THz})^{\frac{1}{2}}}\hspace{-0.5mm}\bigg)^{\hspace{-1mm}s_6} \frac{(\Gamma(-\rho+\hspace{-0.5mm}2\hspace{-0.5mm}+\hspace{-0.5mm}2s_5+s_6))^2}{(\Gamma(-\rho+\hspace{-0.5mm}3\hspace{-0.5mm}+\hspace{-0.5mm}2s_5+s_6))^2} \nonumber \\ & \times  s^{-\frac{1}{2}-s_5-\frac{s_6}{2}} \Gamma(\frac{1}{2}+s_5+\frac{s_6}{2}) ds_5 ds_6
\end{flalign}

Substituting the MGFs of \eqref{eq:mgf_mW}, \eqref{eq:mgf_fso}, and \eqref{eq:mgf_thz} into \eqref{eq:mgf_mrc_eqn} and applying the inverse Laplace transform, we get the PDF of hybrid system for MRC diversity combining as
\begin{flalign} \label{eq:mgf_der_1}
	&f_{\gamma}^{\rm MRC} (\gamma)  = \frac{{{m^m}}}{{\Gamma (m)}}\sum \limits_{j = 0}^\infty {\frac{{K^jd_j }}{{(\Gamma (j + 1))^2(2\zeta ^2)^{j+1}} ({{\bar{\gamma}}^{\rm mW}})^{j+1}}}  \nonumber \\  &\times \frac{1}{2\pi \J} \int_{{L}_3} \Gamma(0-s_3) \Big(\frac{1}{2\zeta ^2 {\bar{\gamma}}^{\rm mW}}\Big)^{s_3} ds_3 \nonumber \\  &\times \Gamma(1+j+s_3)   \frac{\alpha\phi}{\psi\Gamma(\alpha)\Gamma(\beta)} \nonumber \\ & \times \frac{1}{2\pi \J} \int_{{L}_4} \frac{\Gamma(\beta_{\scriptscriptstyle F}-1-s_4)\Gamma(\phi-1-s_4)\Gamma(1+\beta_{\scriptscriptstyle F}+s_4)}{\Gamma(\phi-s_4)} \nonumber \\ & \times \bigg(\frac{\alpha }{\psi ({\bar{\gamma}}^{\rm FSO})^{\frac{1}{2}}}\bigg)^{s_4} ds_4  \Gamma\big(\frac{1}{2}+\frac{ s_4}{2}\big) \sum_{i=1}^{K} \frac{\rho^2 w_i \exp\Big(\frac{-\mu_i^2}{2 \sigma_i^2}\Big) }{2\sqrt{2\pi}\sigma_i ({\bar{\gamma}}^{\rm THz})^{\frac{1}{2}}} \nonumber \\ & \times   \bigg(\frac{1}{2\pi \J}\bigg)^2 \int_{\mathcal{L}_5} \int_{\mathcal{L}_6} \Gamma(0-s_5) \bigg(\frac{1}{{\bar{\gamma}}^{\rm THz}}\bigg)^{s_5}   \Gamma(0-s_6) \nonumber \\ & \times      \bigg(\frac{-2\mu_i}{({\bar{\gamma}}^{\rm THz})^{\frac{1}{2}}}\bigg)^{s_6} \frac{(\Gamma(-\rho+2+2s_5+s_6))^2}{(\Gamma(-\rho+3+2s_5+s_6))^2} ds_5 ds_6 \times  I_3
\end{flalign}

The inner integral $I_3$ can be simplified by substituting $s\gamma = -t$ to get \cite[8.315]{Gradshteyn}
\begin{flalign} \label{eq:mgf_der_2}
&I_3 = \int_{{L}_5} e^{s\gamma} s^{-1-j-s_3} s^{-\frac{1}{2}-\frac{ s_4}{2}} s^{-\frac{1}{2}-s_5-\frac{s_6}{2}}   \nonumber \\ & = \Big(\frac{1}{\gamma}\Big)^{(-1-j -s_3  - \frac{s_4}{2}-s_5-\frac{s_6}{2})} \hspace{-1mm} \frac{2\pi \J}{\Gamma(1+ \hspace{-0.5mm} j \hspace{-0.5mm} +s_3  + \frac{s_4}{2}+s_5+\frac{s_6}{2})}
\end{flalign}

Substituting \eqref{eq:mgf_der_2} into \eqref{eq:mgf_der_1} and using the definition of multivariate Fox's H-function \cite{Mathai_2010}, we get the PDF of hybrid BHL-MRC in \eqref{eq:pdf_hybrid_mrc}. The CDF can be derived using \eqref{eq:pdf_hybrid_mrc} in $F_{\gamma}^{\rm MRC} (\gamma) = \int_{0}^{\gamma} f_{\gamma}^{\rm MRC} (\gamma) d\gamma$. Applying the similar procedure as that of PDF, we get the inner integral as
\begin{flalign}
&\int_{0}^{\gamma} \gamma^{(1+j +s_3+\frac{s_4}{2}+s_5+\frac{s_6}{2}  )} d\gamma \nonumber \\ & = \frac{ \gamma^{(2+j +s_3+\frac{s_4}{2}+s_5+\frac{s_6}{2}  )} \Gamma\big(2+j +s_3+\frac{s_4}{2}+s_5+\frac{s_6}{2}  \big)}{\Gamma\big(3+j +s_3+\frac{s_4}{2}+s_5+\frac{s_6}{2}  \big)}
\end{flalign}

Applying the definition of multivariate Fox's H-function, we get the CDF of hybrid BHL-MRC in \eqref{eq:cdf_hybrid_mrc} to finish the proof of Theorem 2.

\section*{Appendix C}
The average BER of hybrid BHL-MRC can be derived by substituting the CDF of \eqref{eq:cdf_hybrid_mrc} in \eqref{eq:ber_general_eqn} to get
\begin{flalign} \label{eq:ber_hybrid_der}
	&\bar{P}_e^{\rm MRC} = \frac{m^m q^p}{2\Gamma(m) \Gamma(p)}\sum \limits_{j = 0}^\infty {\frac{K^jd_j }{{(\Gamma (j + 1))^2(2\zeta ^2)^{j+1}} ({{\bar{\gamma}}^{\rm mW}})^{j+1}}}  \nonumber \\  &\times \frac{1}{2\pi \J} \int_{{L}_3} \Gamma(0-s_3) \Big(\frac{1}{2\zeta ^2 {\bar{\gamma}}^{\rm mW}}\Big)^{s_3} ds_3 \nonumber \\  &\times \Gamma(1+j+s_3)   \frac{\alpha\phi}{\psi\Gamma(\alpha)\Gamma(\beta)} \nonumber \\ & \times \frac{1}{2\pi \J} \int_{{L}_4} \frac{\Gamma(\beta_{\scriptscriptstyle F}-1-s_4)\Gamma(\phi-1-s_4)\Gamma(1+\beta_{\scriptscriptstyle F}+s_4)}{\Gamma(\phi-s_4)} \nonumber \\ & \times \bigg(\frac{\alpha }{\psi ({\bar{\gamma}}^{\rm FSO})^{\frac{1}{2}}}\bigg)^{s_4} ds_4 \Gamma\big(\frac{1}{2}+\frac{ s_4}{2}\big) \sum_{i=1}^{K} \frac{\rho^2 w_i \exp\Big(\frac{-\mu_i^2}{2 \sigma_i^2}\Big) }{2\sqrt{2\pi}\sigma_i ({\bar{\gamma}}^{\rm THz})^{\frac{1}{2}}} \nonumber \\ & \times   \bigg(\frac{1}{2\pi \J}\bigg)^2 \int_{\mathcal{L}_5} \int_{\mathcal{L}_6} \Gamma(0-s_5) \bigg(\frac{1}{{\bar{\gamma}}^{\rm THz}}\bigg)^{s_5} \nonumber \\   &  \times     \Gamma(0-s_6) \bigg(\frac{-2\mu_i }{({\bar{\gamma}}^{\rm THz})^{\frac{1}{2}}}\bigg)^{s_6} \frac{(\Gamma(-\rho+2+2s_5+s_6))^2}{(\Gamma(-\rho+3+2s_5+s_6))^2} \nonumber \\  & \times   \Gamma(\frac{1}{2}+s_5+\frac{s_6}{2})  \frac{1}{\Gamma(1+j +s_3  + \frac{s_4}{2}+s_5+\frac{s_6}{2})}   \nonumber \\  & \times \frac{ \Gamma\big(2+j +s_3+\frac{s_4}{2}+s_5+\frac{s_6}{2}  \big)}{\Gamma\big(3+j +s_3+\frac{s_4}{2}+s_5+\frac{s_6}{2}  \big)} ds_5ds_6 \times I_4 
\end{flalign}

The inner integral $I_4$ can be solved to \cite[3.381/4]{Gradshteyn}
\begin{flalign} \label{eq:ber_inner}
	I_4 &= \int_{0}^{\infty} e^{-q\gamma}  \gamma^{(1+p+j +s_3+\frac{s_4}{2}+s_5+\frac{s_6}{2}  )}  d\gamma  \nonumber \\   &= q^{(-p-j -s_3-\frac{s_4}{2}-s_5-\frac{s_6}{2})} \Gamma(2\hspace{-0.5mm}+\hspace{-0.5mm}p\hspace{-0.5mm}+\hspace{-0.5mm}j \hspace{-0.5mm} +\hspace{-0.5mm}s_3\hspace{-0.5mm}+\hspace{-0.5mm}\frac{s_4}{2}\hspace{-0.5mm}+\hspace{-0.5mm}s_5\hspace{-0.5mm}+\hspace{-0.5mm}\frac{s_6}{2})
\end{flalign}
substituting the inner integral \eqref{eq:ber_inner} in \eqref{eq:ber_hybrid_der}, and applying the definition of Fox's H-function, we get the average BER of the BHL-MRC in \eqref{eq:ber_hybrid_mrc}. To derive the average BER of the IoT device with maximum SNR in the  access link, we substitute \eqref{eq:cdf_iot} in \eqref{eq:ber_general_eqn} to get
\begin{flalign} \label{eq:ber_mu_der1}
	\bar{P}_e^{\rm IoT} = \frac{q^p}{2\Gamma(p)} \sum_{k=0}^{n} {n \choose k}  \int_{0}^{\infty}  e^{-q\gamma} \gamma^{p-1}     \bigg(-e^{-\frac{\lambda \gamma }{{\bar{\gamma}}^{\rm RF}}}\bigg)^k d\gamma
\end{flalign}

Using the identity \cite[3.381/4]{Gradshteyn}, we get the average BER of the IoT network in \eqref{eq:ber_mu}. Finally, plugging \eqref{eq:ber_hybrid_mrc} and \eqref{eq:ber_mu} into \eqref{eq:ber_relay_eqn}, we get the average BER of the relayed system to finish the proof.

To derive the average BER of BHL-OSC, we substitute \eqref{eq:cdf_sc_eqn} in \eqref{eq:ber_general_eqn} to get
\begin{flalign} \label{eq:ber_osc_der}
	&\bar{P}_e^{\rm OSC} = \frac{q^p}{2\Gamma(p)}  \frac{{{m^m}}}{{\Gamma (m)}}\sum \limits _{j = 0}^\infty {\frac{{K^jd_j }}{{(\Gamma (j + 1))^2}({\bar{\gamma}}^{\rm mW})^j}}   \nonumber \\ & \times   \frac{1}{2\pi \J}\int_{{L}_3} \frac{\Gamma(1+j-s_3) \Gamma(0+s_3)}{\Gamma(1+s_3)} \bigg(\frac{1}{2 \zeta^2 {\bar{\gamma}}^{\rm mW}}\bigg)^{s_3} ds_3 \nonumber \\ & \times  \frac{\alpha_{{\scriptscriptstyle  F}}\phi }{{({\bar{\gamma}}^{\rm FSO})}^{\frac{1}{2}} \psi\Gamma(\alpha)\Gamma(\beta) }  \nonumber \\ & \times  \frac{1}{2\pi \J}\int_{{L}_4} \frac{\Gamma(\beta-1-s_4)\Gamma(\phi-1-s_4)\Gamma(1+\beta+s_4)}{\Gamma(\phi-s_4)\Gamma(2+s_4)}  \nonumber \\ & \times  \Gamma(1+s_4) \bigg(\frac{\alpha }{\psi ({\bar{\gamma}}^{\rm FSO})^{\frac{1}{2}}}\bigg)^{s_4} ds_4 \sum_{i=1}^{K} \frac{\rho^2 w_i \exp\Big(\frac{-\mu_i^2}{2 \sigma_i^2}\Big)}{2 \sqrt{2\pi}\sigma_i (\bar{\gamma}^{\rm THz})^{\frac{1}{2}}}   \nonumber \\ & \times     \bigg(\frac{1}{2\pi \J}\bigg)^2 \int_{\mathcal{L}_5} \int_{\mathcal{L}_6} \Gamma(0-s_5) \bigg(\frac{{1}}{\bar{\gamma}^{\rm THz}}\bigg)^{s_5}   \Gamma(0-s_6) \nonumber \\ & \times     \bigg(\frac{-2\mu_i}{(\bar{\gamma}^{\rm THz})^{\frac{1}{2}}}\bigg)^{s_6} \frac{(\Gamma(-\rho+2+2s_5+s_6))^2}{(\Gamma(-\rho+3+2s_5+s_6))^2}  \nonumber \\ & \times  \frac{\Gamma(2s_5+s_6+1)}{\Gamma(2s_5+s_6+2)}   ds_5 ds_6 \times I_5
\end{flalign}
where $I_5 $ is the inner integral, which is simplified to
\begin{flalign} \label{eq:inner_5}
	I_5 & = \int_{0}^{\infty} e^{-q\gamma} \gamma^{s_3+\frac{s_4}{2}+s_5+\frac{s_6}{2}+j+p}  d\gamma \nonumber \\  & = q^{-1-s_3-\frac{s_4}{2}-s_5-\frac{s_6}{2}-j-p} \Gamma\big(1\hspace{-0.5mm}+\hspace{-0.5mm}s_3\hspace{-0.5mm}+\hspace{-0.5mm}\frac{s_4}{2}\hspace{-0.5mm}+\hspace{-0.5mm}s_5\hspace{-0.5mm}+\hspace{-0.5mm}\frac{s_6}{2}\hspace{-0.5mm}+\hspace{-0.5mm}j\hspace{-1mm}+\hspace{-0.5mm}p\big)  
\end{flalign}
substituting \eqref{eq:inner_5} into \eqref{eq:ber_osc_der}, and applying the definition of multivariate Fox's H-function, we gat the average BER of BHL-OSC as
\begin{flalign}
	&\bar{P}_e^{\rm OSC} = \frac{m^m q^{-1}}{2{\Gamma (m)} \Gamma(p)}\sum \limits _{j = 0}^\infty {\frac{{K^jd_j }}{{(\Gamma (j + 1))^2}({\bar{\gamma}}^{\rm mW})^j}} \nonumber \\  \times & \sum_{i=1}^{K} \frac{\rho^2 w_i \exp\Big(\frac{-\mu_i^2}{2 \sigma_i^2}\Big)}{2 \sqrt{2\pi}\sigma_i (\bar{\gamma}^{\rm THz})^{\frac{1}{2}}} \frac{\alpha_{{\scriptscriptstyle  F}}\phi }{{({\bar{\gamma}}^{\rm FSO})}^{\frac{1}{2}} \psi\Gamma(\alpha)\Gamma(\beta) }  \nonumber \\  \times & H^{0,4:1,1;2,2;0,1;0,1}_{4,3:1,2;3,3;1,0;1,0}\hspace{-0.5mm}\Bigg[\hspace{-0.5mm}\begin{matrix}\frac{\zeta^{-2}\gamma}{2 q  {\bar{\gamma}}^{\rm mW}}, \frac{\alpha \gamma^{\frac{1}{2}} }{\psi (q{\bar{\gamma}}^{\rm FSO})^{\frac{1}{2}}}, \frac{q^{-1}\gamma}{\bar{\gamma}^{\rm THz}}, \frac{-2\mu_i \gamma^{\frac{1}{2}}}{(q\bar{\gamma}^{\rm THz})^{\frac{1}{2}}} \end{matrix} \Bigg| \begin{matrix}  T_9 \\ T_{10}	\end{matrix}\Bigg]
\end{flalign}
where $T_9 = \bigl\{(\rho-1;0,0,2,1)^2, (0;0,0,2,1), (-p-j;1,\frac{1}{2},1,\frac{1}{2}) \bigr\} : \bigl\{ (1,1) \bigr\} ; \bigl\{(-\beta,1), (0,1), (\phi,1)  \bigr\} ; \bigl\{- \bigr\} ; \bigl\{ - \bigr\} $ and $T_10 = \bigl\{(\rho-2;0,0,2,1)^2, (-1;0,0,2,1), (0;0,02,1),  \bigr\} : \bigl\{(1+j,1),(0,1)  \bigr\} ; \bigl\{(\beta-1,1), (\phi-1), (-1,1)   \bigr\} ; \bigl\{(0,1) \bigr\} ; \bigl\{(0,1) \bigr\}$.

\bibliographystyle{ieeetran}
\bibliography{IoT_bib_file}

\begin{thebibliography}{10}
\providecommand{\url}[1]{#1}
\csname url@samestyle\endcsname
\providecommand{\newblock}{\relax}
\providecommand{\bibinfo}[2]{#2}
\providecommand{\BIBentrySTDinterwordspacing}{\spaceskip=0pt\relax}
\providecommand{\BIBentryALTinterwordstretchfactor}{4}
\providecommand{\BIBentryALTinterwordspacing}{\spaceskip=\fontdimen2\font plus
\BIBentryALTinterwordstretchfactor\fontdimen3\font minus
  \fontdimen4\font\relax}
\providecommand{\BIBforeignlanguage}[2]{{%
\expandafter\ifx\csname l@#1\endcsname\relax
\typeout{** WARNING: IEEEtran.bst: No hyphenation pattern has been}%
\typeout{** loaded for the language `#1'. Using the pattern for}%
\typeout{** the default language instead.}%
\else
\language=\csname l@#1\endcsname
\fi
#2}}
\providecommand{\BIBdecl}{\relax}
\BIBdecl

\bibitem{Fuqaha2015_iot_survey}
A.~Al-Fuqaha \emph{et~al.}, ``Internet of things: A survey on enabling
  technologies, protocols, and applications,'' \emph{IEEE Commun. Surv. Tut.},
  vol.~17, no.~4, pp. 2347--2376, 2015.

\bibitem{Luong2016_iot_data_collection_survey}
N.~C. Luong \emph{et~al.}, ``Data collection and wireless communication in
  internet of things ({IoT}) using economic analysis and pricing models: A
  survey,'' \emph{IEEE Commun. Surv. Tut.}, vol.~18, no.~4, pp. 2546--2590,
  2016.

\bibitem{Xu2018_iot_data_collection}
G.~Xu \emph{et~al.}, ``Ubiquitous transmission of multimedia sensor data in
  internet of things,'' \emph{IEEE Internet Things J.}, vol.~5, no.~1, pp.
  403--414, 2018.

\bibitem{Tao2019_iot_data_collection}
H.~Tao \emph{et~al.}, ``Secured data collection with hardware-based ciphers for
  {IoT}-based healthcare,'' \emph{IEEE Internet Things J.}, vol.~6, no.~1, pp.
  410--420, 2019.

\bibitem{Liu2018_iot_data_collection}
S.~Liu \emph{et~al.}, ``Performance analysis of {UAVs} assisted data collection
  in wireless sensor network,'' in \emph{2018 IEEE 87th Veh. Technol. Conf.
  (VTC Spring)}, 2018, pp. 1--5.

\bibitem{Wei2022_iot_data_collection}
Z.~Wei \emph{et~al.}, ``{UAV}-assisted data collection for internet of things:
  A survey,'' \emph{IEEE Internet Things J.}, vol.~9, no.~17, pp.
  15\,460--15\,483, 2022.

\bibitem{Li2023_iot_data_collection}
Y.~Li \emph{et~al.}, ``Data collection maximization in {IoT}-sensor networks
  via an energy-constrained {UAV},'' \emph{IEEE Trans. Mobile Comput.},
  vol.~22, no.~1, pp. 159--174, 2023.

\bibitem{Pang_2014_data_collection}
Y.~Pang \emph{et~al.}, ``Efficient data collection for wireless rechargeable
  sensor clusters in harsh terrains using {UAVs},'' in \emph{2014 IEEE Global
  Commun. Conf.}, 2014, pp. 234--239.

\bibitem{Dehos2014_mW_backhaul}
C.~Dehos \emph{et~al.}, ``Millimeter-wave access and backhauling: the solution
  to the exponential data traffic increase in {5G} mobile communications
  systems?'' \emph{IEEE Commun. Mag.}, vol.~52, no.~9, pp. 88--95, 2014.

\bibitem{Alzenad208_fso_backhaul}
M.~Alzenad \emph{et~al.}, ``{FSO}-based vertical backhaul/fronthaul framework
  for {5G+} wireless networks,'' \emph{IEEE Commun. Mag.}, vol.~56, no.~1, pp.
  218--224, 2018.

\bibitem{Trinh2017_dual_hop_FSO_RF}
P.~V. Trinh \emph{et~al.}, ``Mixed {mmWave} {RF/FSO} relaying systems over
  generalized fading channels with pointing errors,'' \emph{IEEE Photon. J.},
  vol.~9, no.~1, pp. 1--14, 2017.

\bibitem{Li2019_dual_hop_FSO_RF}
R.~Li \emph{et~al.}, ``Performance analysis of a multiuser dual-hop
  amplify-and-forward relay system with {FSO/RF} links,'' \emph{J. Opt. Commun.
  Netw.}, vol.~11, no.~7, pp. 362--370, 2019.

\bibitem{Balti2018_dual_hop_rf_fso}
E.~Balti and M.~Guizani, ``Mixed {RF/FSO} cooperative relaying systems with
  co-channel interference,'' \emph{IEEE Trans. Commun.}, vol.~PP, pp. 1--1, 03
  2018.

\bibitem{Zhang2020_dual_hop_mw_fso}
Y.~Zhang \emph{et~al.}, ``On the performance of dual-hop systems over mixed
  {FSO/mmWave} fading channels,'' \emph{IEEE Open J. Commun. Soc.}, vol.~1, pp.
  477--489, 2020.

\bibitem{Li2022_dual_hop_thz_fso}
S.~Li \emph{et~al.}, ``Mixed {THz/FSO} relaying systems: Statistical analysis
  and performance evaluation,'' \emph{IEEE Trans. Wireless Commun.}, vol.~PP,
  pp. 1--1, 12 2022.

\bibitem{Pai2021_dual_hop_THz_backhaul}
V.~U. Pai \emph{et~al.}, ``Performance analysis of dual-hop {THz} wireless
  transmission for backhaul applications,'' in \emph{2021 IEEE Int. Conf. Adv.
  Netw. Telecommun. Syst. (ANTS)}, 2021, pp. 438--443.

\bibitem{Li_2021_THz_AF}
S.~Li and L.~Yang, ``Performance analysis of dual-hop {THz} transmission
  systems over $\alpha$-$\mu $ fading channels with pointing errors,''
  \emph{IEEE Internet Things J.}, pp. 1--1, 2021.

\bibitem{Bhardwaj2022_systems_journal}
P.~Bhardwaj and S.~M. Zafaruddin, ``Performance of hybrid {THz} and
  multiantenna {RF} system with diversity combining,'' \emph{IEEE Syst. J.},
  pp. 1--12, 2022.

\bibitem{Chen2016_hybrid_rf_fso}
L.~Chen \emph{et~al.}, ``Multiuser diversity over parallel and hybrid {FSO/RF}
  links and its performance analysis,'' \emph{IEEE Photon. J.}, vol.~8, no.~3,
  pp. 1--9, 2016.

\bibitem{Makki2016_hybrid_fso_RF}
B.~Makki \emph{et~al.}, ``On the performance of {RF-FSO} links with and without
  hybrid {ARQ},'' \emph{IEEE Trans. Wireless Commun.}, vol.~15, no.~7, pp.
  4928--4943, 2016.

\bibitem{Dahrouj2015_rf_fso_hybrid_backhaul}
H.~Dahrouj \emph{et~al.}, ``Cost-effective hybrid {RF/FSO} backhaul solution
  for next generation wireless systems,'' \emph{IEEE Wireless Commun.},
  vol.~22, no.~5, pp. 98--104, 2015.

\bibitem{Enayati2016_hybrid_fso_RF_backhaul}
S.~Enayati and H.~Saeedi, ``Deployment of hybrid {FSO/RF} links in backhaul of
  relay-based rural area cellular networks: Advantages and performance
  analysis,'' \emph{IEEE Commun. Lett.}, vol.~20, no.~9, pp. 1824--1827, 2016.

\bibitem{Eryani2018_hybrid_rf_fso}
Y.~F. Al-Eryani \emph{et~al.}, ``Protocol design and performance analysis of
  multiuser mixed {RF} and hybrid {FSO/RF} relaying with buffers,'' \emph{J.
  Opt. Commun. Netw.}, vol.~10, no.~4, pp. 309--321, 2018.

\bibitem{Sharma2019_hybrid_rf_fso}
S.~Sharma \emph{et~al.}, ``Switching-based cooperative decode-and-forward
  relaying for hybrid {FSO/RF} networks,'' \emph{J. Opt. Commun. Netw.},
  vol.~11, no.~6, pp. 267--281, 2019.

\bibitem{Badarneh2020_mW_FSO_simultaneous}
O.~S. Badarneh and R.~Mesleh, ``Diversity analysis of simultaneous mmwave and
  free-space-optical transmission over $\cal{F}$-distribution channel models,''
  \emph{J. Opt. Commun. Netw.}, vol.~12, no.~11, pp. 324--334, 2020.

\bibitem{Singya2022_hybrid_fso_thz_backhaul}
P.~K. Singya \emph{et~al.}, ``Hybrid {FSO/THz}-based backhaul network for
  {mmWave} terrestrial communication,'' \emph{IEEE Trans. Wireless Commun.},
  pp. 1--1, 2022.

\bibitem{Hasna_2004_AF}
M.~Hasna and M.-S. Alouini, ``A performance study of dual-hop transmissions
  with fixed gain relays,'' \emph{IEEE Trans. Wireless Commun.}, vol.~3, no.~6,
  pp. 1963--1968, 2004.

\bibitem{Mathai_2010}
A.~M.{ Mathai} \emph{et~al.}, ``The {H}-function: Theory and applications.''
  vol. New York, NY, USA, Springer, 2010.

\bibitem{Jiayi2018_FTR}
J.~Zhang \emph{et~al.}, ``New results on the fluctuating two-ray model with
  arbitrary fading parameters and its applications,'' \emph{IEEE Trans. Veh.
  Technol.}, vol.~67, no.~3, pp. 2766--2770, 2018.

\bibitem{Miguel2021}
M.~López-Benítez and J.~Zhang, ``Comments and corrections to “{New} results
  on the fluctuating two-ray model with arbitrary fading parameters and its
  applications”,'' \emph{IEEE Trans. Veh. Technol.}, vol.~70, no.~2, pp.
  1938--1940, 2021.

\bibitem{Najafi2020_uav_pointing}
M.~Najafi \emph{et~al.}, ``Statistical modeling of the {FSO} fronthaul channel
  for {UAV}-based communications,'' \emph{IEEE Trans. Commun.}, vol.~68, no.~6,
  pp. 3720--3736, 2020.

\bibitem{Boulogeorgos_PE_2022}
A.-A.~A. Boulogeorgos \emph{et~al.}, ``Outage performance analysis of
  {RIS}-assisted {UAV} wireless systems under disorientation and
  misalignment,'' \emph{IEEE Trans. Veh. Technol.}, vol.~71, no.~10, pp.
  10\,712--10\,728, 2022.

\bibitem{Papasotiriou2023_scintific_report_outdoor}
E.~Papasotiriou \emph{et~al.}, ``Outdoor {THz} fading modeling by means of
  gaussian and gamma mixture distributions,'' \emph{Scientific Rep.}, vol.~13,
  no. 6385, 2023.

\bibitem{Badarneh2023_THz_Pointing}
O.~S. Badarneh, M.~T. Dabiri, and M.~Hasna, ``Channel modeling and performance
  analysis of directional {THz} links under pointing errors and $\alpha$-$\mu$
  distribution,'' \emph{IEEE Commun. Lett.}, pp. 1--1, 2023.

\bibitem{Bletsas2006}
A.~Bletsas \emph{et~al.}, ``A simple cooperative diversity method based on
  network path selection,'' \emph{IEEE J. Sel. Areas Commun.}, vol.~24, no.~3,
  pp. 659--672, 2006.

\bibitem{papoulis_2002}
{A. Papoulis} and {S. Pillai}, \emph{Probability, Random Variables, and
  Stochastic Processes}.\hskip 1em plus 0.5em minus 0.4em\relax McGraw Hill,
  Boston, Fourth Edition, 2002.

\bibitem{Ansari2011}
I.~S. {Ansari} \emph{et~al.}, ``A new formula for the {BER} of binary
  modulations with dual-branch selection over generalized-k composite fading
  channels,'' \emph{IEEE Trans. Commun.}, vol.~59, no.~10, pp. 2654--2658,
  2011.

\bibitem{Tsiftsis2006}
T.~A. Tsiftsis \emph{et~al.}, ``Multihop free-space optical communications over
  strong turbulence channels,'' in \emph{2006 IEEE Int. Conf. Commun.}, vol.~6,
  2006, pp. 2755--2759.

\bibitem{Boulogeorgos_Analytical}
A.~A. {Boulogeorgos} and A.~{Alexiou}, ``Analytical performance assessment of
  {THz} wireless systems,'' \emph{IEEE Access}, vol.~7, pp. 11\,436--11\,453,
  2019.

\bibitem{Mittal_1972}
P.~{Mittal} and K.~{Gupta}, ``An integral involving generalized function of two
  variables,'' \emph{Proc. Indian Acad. Sci.}, vol.~75, no.~9, pp. 117--123,
  1972.

\bibitem{Gradshteyn}
I.~S. {Gradshteyn} and I.~M. {Ryzhik }, \emph{Table of Integrals, Series, and
  Products}.\hskip 1em plus 0.5em minus 0.4em\relax Academic press, San Diego,
  CA, 6th edition, 2000.

\end{thebibliography}
\end{document}